\documentclass[12pt,reqno]{amsart}

\usepackage{epsfig}
\usepackage{amsmath, amsthm, amsfonts}
\usepackage{graphicx}

\numberwithin{equation}{section}

\newcommand{\di}{\displaystyle}
\newcommand{\adj}{{\rm adj}\,}

\newcommand{\R}{{\mathbb R}}
\newcommand{\C}{{\mathbb C}}

\newcommand{\Z}{{\mathbb Z}}
\newcommand{\T}{{\mathbb T}}

\newcommand{\B}{{\mathcal B}}

\renewcommand{\Re}{{\operatorname{Re\,}}}
\renewcommand{\Im}{{\operatorname{Im\,}}}

\newcommand{\Tr}{{{\operatorname{Tr}\,}}}

\newcommand{\al}{\alpha}
\newcommand{\be}{\beta}
\newcommand{\ga}{\gamma}
\newcommand{\Ga}{\Gamma}
\newcommand{\La}{\Lambda}
\newcommand{\la}{\lambda}
\newcommand{\ep}{\varepsilon}

\newcommand{\f}{\varphi}
\newcommand{\sg}{\sigma}

\newcommand{\om}{\omega}

\newcommand{\bq}{\begin{eqnarray}}
\newcommand{\eq}{\end{eqnarray}}
\newcommand{\nn}{\nonumber}
\newcommand{\ba}{\begin{array}}
\newcommand{\ea}{\end{array}}

\newtheorem{theo}{{\sc \bf Theorem}}[section]

\begin{document}

\title[Dimer model on a triangular lattice]
{Exact solution of the classical dimer model on a triangular lattice: Monomer-monomer correlations}

\author{Estelle Basor}
\address{American Institute of Mathematics,
600 East Brokaw Rd. 
San Jose, CA 95112}
\email{ebasor@aimath.org}

\author{Pavel Bleher}
\address{Department of Mathematical Sciences,
Indiana University-Purdue University Indianapolis,
402 N. Blackford St., Indianapolis, In 46202, U.S.A.}
\email{bleher@math.iupui.edu}

\thanks{The first author is supported in part
by the National Science Foundation (NSF) Grants DMS-1265172 and DMS-1565602.}

\date{}

\begin{abstract}
We obtain an asymptotic formula, as $n\to\infty$, 
for the monomer-monomer correlation function $K_2(x,y)$ in the classical dimer
model on a triangular lattice, with the horizontal and vertical weights $w_h=w_v=1$
and the diagonal weight $w_d=t>0$, where $x$ and $y$ are sites $n$ spaces apart in adjacent rows.
 We find that $t_c=\frac{1}{2}$ is a critical value of $t$. We prove that in the subcritical case, 
$0<t<\frac{1}{2}$, as $n\to\infty$, $K_2(x,y)=K_2(\infty)\left[1-\di\frac{e^{-n/\xi}}{n}\,\Big(C_1+C_2(-1)^n+\mathcal O(n^{-1})\Big)
\right]$, with explicit formulae for $K_2(\infty)$, $\xi$, $C_1$, and $C_2$.
In the supercritical case, $\frac{1}{2} < t < 1$, we prove that as $n\to\infty$, 
$K_2(x,y)=K_2(\infty)\Bigg[1-
\di\frac{e^{-n/\xi}}{n}\,
\Big(C_1\cos(\omega n+\varphi_1)+C_2(-1)^n\cos(\omega n+\varphi_2)+
C_3+C_4(-1)^n$ $+\mathcal O(n^{-1})\Big)\Bigg]$,
with explicit formulae for $K_2(\infty)$, $\xi$, $\om$, and  $C_1$, $C_2$, $C_3$, $C_4$, $\f_1$, $\f_2$.
The proof is based on an extension of the Borodin--Okounkov--Case--Geronimo formula to block Toeplitz determinants and on
an asymptotic analysis of the Fredholm determinants in hand.
\end{abstract}


\maketitle

\section{Introduction}

This work is a continuation of the works of Fendley, Moessner, and Sondhi \cite{FMS} 
and Basor and Ehrhardt \cite{BE}. We consider the classical dimer model on a triangular lattice with the weights 
\begin{equation}\label{intro1}
w_h=w_v=1,\qquad w_d=t>0.
\end{equation}
It is convenient to view the triangular lattice as a square lattice with diagonals, as shown on Fig.1.
\begin{center}
\begin{figure}[h]
\begin{center}
\scalebox{0.55}{\includegraphics{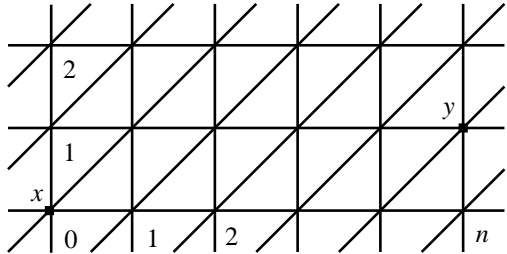}}
\end{center}
  \caption[sectors ]{The triangular lattice as a square lattice with diagonals.}
 \end{figure}
\label{tlat}
\end{center}

Our main goal is to calculate an asymptotic behavior as $n\to\infty$ of the monomer-monomer correlation function $K_2(x,y)$
in the thermodynamic limit,  
 where $x$ and $y$ are sites $n$ spaces apart in adjacent rows (see Fig.1). As explained in \cite{FMS},
this problem has important applications to 
the quantum dimer model on the triangular lattice, for the study of the resonating valence bond (RVB) phase and the 
ground state degeneracies on closed surfaces. In this respect 
see the works on the quantum dimer model and the RVB phase by  Rokhsar and Kivelson \cite{RK}, 
Moessner, Sondhi \cite{MS1,MS2}, and others. See also the lectures by Moessner and Raman \cite{MR} and references therein, and
the review collection \cite{DIEP}.

To derive an asymptotic behavior as $n\to\infty$ of the monomer-monomer correlation function $K_2(x,y)$ we apply the following steps:
\begin{enumerate}
\item We use a representation of  $K_2(x,y)$ as a block Toeplitz determinant. This representation was obtained in the work \cite{BE}
from a different determinantal representation of $K_2(x,y)$ developed in \cite{FMS}.
\item We write the block Toepltiz determinant in terms of a Fredholm determinant through an extension
of  Borodin--Okounkov--Case--Geronimo formula to block Toeplitz determinants as described in \cite{BE}.
\item We calculate explicitly the Wiener--Hopf factorization of the $2\times 2$ matrix symbol under consideration.
To do this we use a {\it power decreasing algorithm}, the idea of which goes back to the work of McCoy and Wu \cite{MW}.
\item We analyze an asymptotic behavior of the Fourier coefficients of the Wiener--Hopf factors.
\item Finally, we obtain the desired asymptotic behavior of the monomer-monomer correlation function $K_2(x,y)$.
The calculations are somewhat different for the subcritical, $t<\frac{1}{2}\,$, and supercritical, $\frac{1}{2} < t < 1\,$, 
cases.
\end{enumerate}

The set-up for the rest of the paper is as follows. In Section \ref{blockT} we  
present a formula for $K_2(x,y)$ in terms of a block Toeplitz determinant. In Section \ref{BOCG} 
and Appendix \ref{Appendix_A} we discuss and prove a  
 Borodin--Okounkov--Case--Geronimo type formula for block Toeplitz determinants.  In Section
\ref{WH} and Appendix \ref{Appendix_B} we calculate the Wiener--Hopf factorization of the $2\times 2$ matrix-valued
symbol in hand. Section \ref{sub_case} is devoted to the asymptotic formula for $K_2(x,y)$ in the subcritical case, $t < \frac{1}{2} \,$,
and Section \ref{super_case} to the one  in the supercritical case, $\frac{1}{2} < t < 1\,$.  In particular, in Section \ref{super_case}
we calculate numerical values for the constants in the asymptotic formula  for $K_2(x,y)$ in the symmetric case for the limiting value at $t=1$.  

It should be mentioned that the technique used to find the factorization for the block-Toeplitz operator, or equivalently the Wiener-Hopf factorization of the symbol, is of independent interest and can be used in other situations. In particular, this method is useful for operators whose symbols are of the form a scalar function times a symbol whose entries are trigonometric polynomials. It requires that the determinant of the symbol be of a ``degree'' that in some sense matches the degrees of the polynomials. This will be clear by the example done in this paper. In these cases, it is surprising that the factorization can really be found by knowledge of the determinant of the symbol alone. 

As mentioned above, this method appears first in the McCoy and Wu paper \cite{MW} where it was used to find asymptotic expansions for the correlations for the two-dimensional Ising model.  The authors are especially grateful for the many useful discussions with Barry McCoy that took place at the Simons Center for Geometry and Physics program ''Statistical mechanics and combinatorics'' held 
from mid-February to mid-April of 2016 and for the generous support of the Center as well.

\section{Monomer-monomer correlation function as a block Toeplitz determinant}\label{blockT}

Our starting point is a determinantal formula for $K_2(x,y)$ obtained in \cite{FMS}. Namely, 
 the monomer-monomer correlation function $K_2(x,y)$ can be expressed in terms 
of the $(2n)\times (2n)$ determinant as  
\begin{equation}\label{intro2}
K_2(x,y)=\frac{1}{2}\sqrt{\det
\begin{pmatrix}
\mathcal R & \mathcal Q \\
\mathcal Q & \mathcal R
\end{pmatrix}},
\end{equation}
where the matrix elements of the $n\times n$ matrices $\mathcal R$ and $\mathcal Q$ are equal to
\begin{equation}\label{intro3}
\begin{aligned}
&\mathcal R_{jk}=2(-1)^{\lfloor\frac{k-j}{2}\rfloor}R_{k-j+1}+\theta(j-k)t^{j-k-1},\\
&\mathcal Q_{jk}=2i(-1)^{\lfloor\frac{j+k}{2}\rfloor}Q_{n+1-k-j},
\end{aligned}
\end{equation}
with $\lfloor k\rfloor $ being the floor function and $R_k$, $Q_k$, $\theta(k)$ defined as follows:
For even $k$,
\begin{equation}\label{intro4}
\begin{aligned}
R_k=\frac{1}{8\pi^2}\int_{-\pi}^\pi\int_{-\pi}^\pi \frac{\cos y\cos (kx+y)\,dxdy}
{\cos^2x+\cos^2y +t^2\cos^2(x+y)}\,,
\end{aligned}
\end{equation}
and $Q_k=0$. For odd $k$,
\begin{equation}\label{intro5}
\begin{aligned}
Q_k=\frac{1}{8\pi^2}\int_{-\pi}^\pi\int_{-\pi}^\pi \frac{\cos x\cos (kx)\,dxdy}
{\cos^2x+\cos^2y +t^2\cos^2(x+y)}\,,
\end{aligned}
\end{equation}
and
\begin{equation}\label{intro6}
\begin{aligned}
R_k=\frac{t}{8\pi^2}\int_{-\pi}^\pi\int_{-\pi}^\pi \frac{\cos (x+y)\cos (kx+y)\,dxdy}
{\cos^2x+\cos^2y +t^2\cos^2(x+y)}\,,
\end{aligned}
\end{equation}
The expression $\theta(k)$ equals 1 for $k>0$ and 0 otherwise. 

As shown in \cite{BE}, formula \eqref{intro2} can be further
reduced to a block Toeplitz determinant. Namely, we have that
\begin{equation}\label{mm2}
K_2(x,y)=\frac{1}{2}\sqrt{\det T_n(\phi)},
\end{equation}
where $T_n(\phi)$ is the finite block Toeplitz matrix,
\begin{equation}\label{mm3}
 T_n(\phi)=(\phi_{j-k}),\quad 0\le j,k\le n-1,
\end{equation}
where
\begin{equation}\label{mm4}
\phi_k=\frac{1}{2\pi}\int_0^{2\pi} \phi(e^{ix})e^{-ikx}dx
\end{equation}
and
\begin{equation}\label{mm5}
 \phi(e^{ix})=\sg(e^{ix})
\begin{pmatrix}
p(e^{ix}) & q(e^{ix}) \\
q(e^{-ix}) & p(e^{-ix})
\end{pmatrix},
\end{equation}
with
\begin{equation}\label{mm6}
\sg(e^{ix})=\frac{1}{(1-2t\cos x+t^2)\sqrt{t^2+\sin^2x+\sin^4 x}}
\end{equation}
and
\begin{equation}\label{mm7}
p(e^{ix})=(t\cos x+\sin^2x)(t-e^{ix}),\qquad 
q(e^{ix})=\sin x(1-2t\cos x+t^2).
\end{equation}
Observe that
\begin{equation}\label{mm8}
\det\phi(z)=\frac{1}{1-2t\cos x+t^2}=\frac{1}{(z-t)(z^{-1}-t)}\,,\qquad z=e^{ix}.
\end{equation}

\section{Borodin--Okounkov--Case--Geronimo type formula} \label{BOCG}

To evaluate the asymptotics of $\det T_n(\phi)$ as $n\to\infty$ we use a 
Borodin--Okounkov--Case--Geronimo (BOCG) type formula
for block Toeplitz determinants. For the original, scalar version of the 
BOCG formula see the papers of Geronimo--Case \cite{GC}, Borodin--Okounkov \cite {BO},
Basor--Widom \cite{BW}, B\"ottcher \cite{Bot}, and references therein. For an asymptotic formula for block
 Toeplitz determinants see the earlier works of Widom \cite{Wid1,Wid2}.

For any matrix-valued $2\pi$-periodic function $\f(e^{ix})$ consider the 
corresponding semi-infinite matrices, Toeplitz and Hankel,
\begin{equation}\label{bo1}
T(\f)=(\f_{j-k})_{j,k=0}^\infty\,; \qquad H(\f)=(\f_{j+k+1})_{j,k=0}^\infty\,, 
\end{equation}
where
\begin{equation}\label{bo2}
\f_k=\frac{1}{2\pi}\int_0^{2\pi} \f(e^{ix})e^{-ikx}dx
\end{equation}
Let
\begin{equation}\label{bo3}
\psi(e^{ix})=\phi^{-1}(e^{ix}),
\end{equation}
where $\phi(e^{ix})$ is as in \eqref{mm5}.
Then the following Borodin--Okounkov--Case--Geronimo type formula holds (see \cite{BE} and Appendix \ref{Appendix_A} below):
\begin{equation}\label{bo4}
\det T_n(\phi)=\frac{E(\psi)}{G(\psi)^n}\,\det\left(I-\Phi\right),
\end{equation}
where $\det\left(I-\Phi\right)$ is the Fredholm determinant  with
\begin{equation}\label{bo5}
\begin{aligned}
\Phi
=H\big(e^{-inx}\psi(e^{ix})\big)T^{-1}\big(\psi(e^{-ix})\big)
H\big(e^{-inx}\psi(e^{-ix})\big)
T^{-1}\big(\psi(e^{ix})\big).
\end{aligned}
\end{equation}
This formula, under some conditions, holds for a general symbol $\phi$, but we will use for $\phi$ as
in \eqref{mm5}. In this case $G(\psi)=1$ and as shown in \cite{BE},
\begin{equation}\label{bo6}
E(\psi)=\frac{t}{2t(2+t^2)+(1+2t^2)\sqrt{2+t^2}}\,.
\end{equation}
In virtue of \eqref{mm2}, this implies that the order parameter is equal to
\begin{equation}\label{bo7}
K_2(\infty)=\lim_{n\to\infty} K_2(x,y)=\frac{1}{2}\sqrt{E(\psi)}\,=\frac{1}{2}\sqrt{\frac{t}{2t(2+t^2)+(1+2t^2)\sqrt{2+t^2}}}\,.
\end{equation}
Our goal is evaluate an asymptotic behavior of $K_2(x,y)$ as $n\to\infty$. By \eqref{mm2} and \eqref{bo4}, the
problem reduces to an asymptotic behavior of the Fredholm determinant $\det\left(I-\Phi\right)$, because 
\begin{equation}\label{bo8}
 K_2(x,y)=K_2(\infty)\sqrt{\det\left(I-\Phi\right)}\,.
\end{equation}

\section{Wiener--Hopf factorization  of $\phi(z)$}\label{WH}

To evaluate $\det\left(I-\Phi\right)$ we use the Wiener--Hopf factorization. To that end we first need
to factor $\phi$.
Let $z=e^{ix}$. Denote
\begin{equation}\label{p1}
\begin{aligned}
 \pi(z)&=
\begin{pmatrix}
p(z) & q(z) \\
q(z^{-1}) & p(z^{-1})
\end{pmatrix},
\end{aligned}
\end{equation}
so that equation \eqref{mm5} reads
\begin{equation}\label{p2}
 \phi(z)=\sg(z)\pi(z),
\end{equation}
where $\sg(z)$ is a scalar function, see \eqref{mm6}.
From \eqref{mm8} and \eqref{mm6} we obtain that
\begin{equation}\label{p3}
\det \pi(z)=(t^2+\sin^2 x+\sin^4 x)(z-t)(z^{-1}-t).
\end{equation}
Furthermore, we have the following factorization (see \cite{BE}):
\begin{equation}\label{p5}
t^2+\sin^2x+\sin^4x=\frac{1}{16 \xi_1\xi_2}
\left(z^{-2}-\xi_1\right)\left( z^{-2}-\xi_2\right)
\left(z^2-\xi_1\right)\left(z^2-\xi_2\right),
\end{equation}
where
\begin{equation}\label{p6}
\xi_{1,2}=2\pm \mu-2\sqrt{1-t^2\pm \mu},\quad \mu=\sqrt{1-4t^2}\,.
\end{equation}
Observe that
\begin{equation}\label{p7}
0<|\xi_{1,2}|<1.
\end{equation}

Denote now
\begin{equation}\label{p8}
\tau=\frac{1}{t}>1\,;\qquad \eta_j=\frac{1}{\sqrt\xi_j}\,,\quad |\eta_j|>1,\quad j=1,2.
\end{equation}
In what follows we use $\tau$ and $\eta_j$, $j=1,2$, as the main parameters. The graphs of $\eta_1(t)$ and $\eta_2(t)$
 are shown on Fig.2. The functions $\eta_1(t)$, $\eta_2(t)$ are positive for $0\le t\le \frac{1}{2}$, 
and complex-valued for $t>\frac{1}{2}$. 
For $t>\frac{1}{2}$ the graph shows $|\eta_1(t)|=|\eta_2(t)|$ and $\arg \eta_1(t)$, $\arg \eta_2(t)$.
\begin{center}
\begin{figure}[h]
\begin{center}
\scalebox{0.45}{\includegraphics{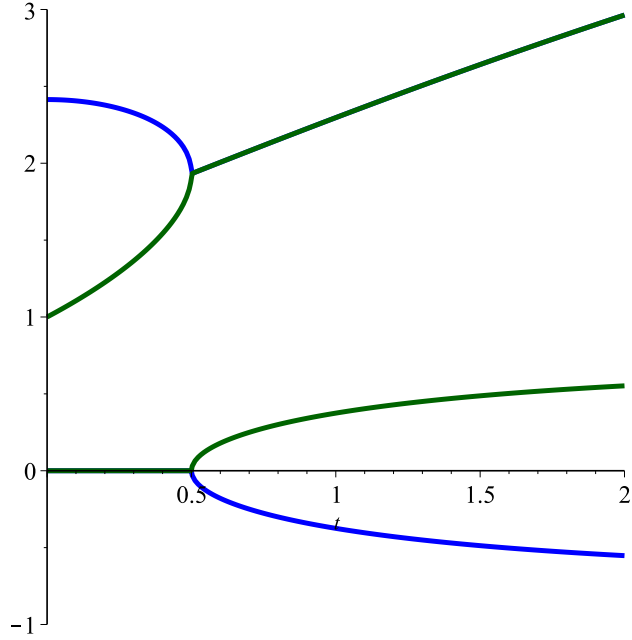}}
\end{center}
  \caption[sectors ]{The graphs of $|\eta_1(t)|$ (in blue), $|\eta_2(t)|$  (in green), the upper graphs,
and $\arg\eta_1(t)$, $\arg\eta_2(t)$, the lower graphs.}
 \end{figure}
\label{eta}
\end{center}
In terms of these parameters, equation \eqref{p5} is written as
\begin{equation}\label{p9}
t^2+\sin^2x+\sin^4x=\frac{1}{16 \eta_1^2\eta_2^2}
\left(z^{-2}-\eta_1^2\right)\left( z^{-2}-\eta_2^2\right)
\left(z^2-\eta_1^2\right)\left(z^2-\eta_2^2\right),
\end{equation}
and \eqref{mm7} as
\begin{equation}\label{p10}
p(z)=\tau^{-2}z(\cos x+\tau\sin^2x)(z^{-1}-\tau),\qquad 
q(z)=\tau^{-2}\sin x\,(z-\tau)(z^{-1}-\tau),
\end{equation}
hence by \eqref{p1},
\begin{equation}\label{p11}
\begin{aligned}
 \pi(z)&=
\tau^{-2}\begin{pmatrix}
z(\cos x+\tau\sin^2x)(z^{-1}-\tau) & \sin x\,(z-\tau)(z^{-1}-\tau) \\
-\sin x\,(z-\tau)(z^{-1}-\tau) & z^{-1}(\cos x+\tau\sin^2x)(z-\tau)
\end{pmatrix}.
\end{aligned}
\end{equation}
By \eqref{p3},
\begin{equation}\label{p11a}
\det \pi(z)=\tau^{-2}(t^2+\sin^2 x+\sin^4 x)(z-\tau)(z^{-1}-\tau),
\end{equation}
hence by \eqref{p9},
\begin{equation}\label{p11b}
\det \pi(z)=g(z)g(z^{-1}),
\end{equation}
where
\begin{equation}\label{p11c}
g(z)=\frac{(z-\tau)(z^2-\eta_1^2)(z^2-\eta_2^2)}{4\tau\eta_1\eta_2}\,.
\end{equation}
Observe that $g(z)$ does not vanish on the unit disk $D=\{|z|\le 1\}$.

{\it Symmetry relations.} From \eqref{mm5}-\eqref{mm7},
\begin{equation}\label{sym1}
\tilde\phi(e^{ix})\equiv\phi(e^{-ix})=\adj\phi(e^{ix}),\quad
{\rm adj}\begin{pmatrix}
a & b \\
c & d
\end{pmatrix}
\equiv\begin{pmatrix}
d & -b \\
-c & a
\end{pmatrix},
\end{equation}
and
\begin{equation}\label{sym2}
\tilde\psi(e^{ix})\equiv\psi(e^{-ix})=\phi^{-1}(e^{-ix})=\adj  \psi(e^{ix}).
\end{equation}
The matrix valued function $\pi(z)$ satisfies the symmetry relation,
\begin{equation}\label{sym2a}
\pi\left(z^{-1}\right)=\adj \pi(z)=\pi^{-1}(z)\det \pi(z),
\end{equation}
or equivalently, by \eqref{p11b},
\begin{equation}\label{sym3}
 \left[\frac{\pi(z)}{g(z)}\right]^{-1}=\frac{\pi\left(z^{-1}\right)}{g(z^{-1})}\,.
\end{equation}
We will use this relation in a factorization of $\pi(z)$. 

The function $\phi$ in \eqref{mm5} also satisfies the relation
\begin{equation}\label{sym4}
 \phi^{\rm T}(z)=\sg_3\phi(z)\sg_3,\quad \sg_3=
\begin{pmatrix}
1 & 0 \\
0 & -1
\end{pmatrix},
\end{equation}
where $\phi^{\rm T}$ is the transpose of the matrix $\phi$.

{\it Equation in $\eta_1,\eta_2$.} The numbers $\eta_1,\eta_2$ in \eqref{p8} satisfy the equation
\begin{equation}\label{p13}
(\eta_1^2-1)(\eta_2^2-1)=\frac{4\eta_1\eta_2}{\tau}\,.
\end{equation} 

Our goal is to factor $\phi(z)$ as 
$\phi(z)= \phi_+(z)\phi_-(z)$, 
where $\phi_+(z)$ and $\phi_-(z^{-1})$ are analytic invertible matrix valued functions on the disk $D=\{z\;|\;|z|\le 1\}$.

\begin{theo}\label{factor} We have the factorization:
\begin{equation}\label{th1}
\phi(z)= \phi_+(z)\phi_-(z),
\end{equation}
where 
\begin{equation}\label{th2}
\phi_+(z)=A(z)\Psi(z),\quad \phi_-(z)=\Psi^{-1}(z^{-1}),
\end{equation}
with
\begin{equation}\label{th3}
\begin{aligned}
A(z)&=\frac{\tau}{z-\tau}\,,
\end{aligned}
\end{equation}
and
\begin{equation}\label{th4}
\Psi(z)=\frac{1}{\sqrt{f(z)}}\,D_0(z)P_1D_1(z)P_2D_2(z)P_3D_3(z)P_4D_4(z)P_5,
\end{equation}
with
\begin{equation}\label{th4a}
f(z)=\frac{(z^2-\eta_1^2)(z^2-\eta_2^2)}{4\eta_1\eta_2}\,
\end{equation}
and
\begin{equation}\label{th5}
\begin{aligned}
D_0(z)&=
\begin{pmatrix}
1 & 0 \\
0 & z-\tau
\end{pmatrix}, 
\quad 
D_1(z)=\begin{pmatrix}
z-\eta_1 & 0 \\
0 & 1
\end{pmatrix},\quad
D_2(z)=\begin{pmatrix}
z+\eta_1 & 0 \\
0 & 1
\end{pmatrix},\\
D_3(z)&=\begin{pmatrix}
z-\eta_2 & 0 \\
0 & 1
\end{pmatrix},\quad
D_4(z)=\begin{pmatrix}
1 & 0 \\
0 & z+\eta_2
\end{pmatrix},
\end{aligned}
\end{equation}
and
\begin{equation}\label{th6}
\begin{aligned}
P_j=\begin{pmatrix}
1 & p_j \\
0 & 1
\end{pmatrix},\quad j=1,2,3,5;\quad
P_4=\begin{pmatrix}
1 & 0 \\
p_4 & 1
\end{pmatrix}.
\end{aligned}
\end{equation}
Here
\begin{equation}\label{th7}
\begin{aligned}
p_1&=\frac{i\big[\tau(\eta_1^2-1)^2-2\eta_1(\eta_1^2+1)\big]}{2(\eta_1^2-1)}\,,\quad
p_2=-\frac{i(\eta_1^2+1)}{\eta_1^2-1}\,,\\
p_3&=\frac{i\tau(\eta_1+1)}{2\eta_1}\,,\quad
p_4=-\frac{2i\eta_1\eta_2}{\tau}\,,\quad p_5=-\frac{i\tau}{2\eta_1}\,.
\end{aligned}
\end{equation}
\end{theo}

{\it Remark.} The last factor, $P_5$, in \eqref{th4} cancels out in factorization \eqref{th1}, so it can be
any constant invertible matrix. Our choice of $P_5$ in \eqref{th6}, \eqref{th7} ensures a symmetry of $\Psi(z)$ 
with respect to the swap of $\eta_1$, $\eta_2$. See formulae \eqref{fd4}, \eqref{fd4a} below.
 This symmetry will be important in the subsequent calculations.  

A proof of Theorem \ref{factor} is given in Appendix \ref{Appendix_B} below. Applying symmetry relation \eqref{sym4} to
\eqref{th1}, we obtain a minus-plus factorization of $\phi(z)$:
\begin{equation}\label{th8}
\phi(z)= \sg_3\phi^{\rm T}(z)\sg_3= \sg_3[\phi_+(z)\phi_-(z)]^{\rm T}\sg_3
=[\sg_3 \phi_-^{\rm T}(z)]\,[\phi_+^{\rm T}(z)\sg_3],
\end{equation}
hence
\begin{equation}\label{th9}
\phi(z)= \theta_-(z)\theta_+(z),
\end{equation}
where 
\begin{equation}\label{th10}
\theta_-(z)=\sg_3 \phi_-^{\rm T}(z),\quad \theta_+(z)=\phi_+^{\rm T}(z)\sg_3.
\end{equation}

\section{Asymptotics of the monomer-monomer correlation function. Subcritical case, $0<t<\frac{1}{2}$} \label{sub_case}

We have another useful representation of $\det(1-\Phi)$ (see \cite{BE} and Appendix \ref{Appendix_A} below): 
\begin{equation}\label{fdet1}
\det(I-\Phi)=\det(I-\La),
\end{equation}
where
\begin{equation}\label{fdet2}
\La=H(z^{-n}\al)H( z^{-n}\be)
\end{equation}
with
\begin{equation}\label{fdet3}
\al(z)=\phi_-(z)\theta_+^{-1}(z),\quad \be(z)=\theta_-^{-1}(z^{-1})\phi_+(z^{-1}).
\end{equation}
The matrix elements of the matrix $\La $ are 
\begin{equation}\label{fd1}
\begin{aligned}
\La_{jk}
=\sum_{a=0}^\infty \al_{j+n+a+1}\be_{k+n+a+1},
\end{aligned}
\end{equation}
where
\begin{equation}\label{fd2}
\begin{aligned}
\al_k=\frac{1}{2\pi}\int_0^{2\pi} \al(e^{ix})e^{-ikx}dx,\quad
\be_k=\frac{1}{2\pi}\int_0^{2\pi} \be(e^{ix})e^{-ikx}dx.
\end{aligned}
\end{equation}
Let us calculate an asymptotic behavior of $\al_k$, $\be_k$ as $k\to \infty$.
Denote 
\begin{equation}\label{fd3}
\begin{aligned}
D(z)=D_0(z)P_1D_1(z)P_2D_2(z)P_3D_3(z)P_4D_4(z)P_5,
\end{aligned}
\end{equation}
so that by \eqref{th4},
\begin{equation}\label{fd4}
\begin{aligned}
\Psi(z)=\frac{1}{\sqrt{f(z)}}\,D(z),\quad f(z)=\frac{(z^2-\eta_1^2)(z^2-\eta_2^2)}{4\eta_1\eta_2}\,.
\end{aligned}
\end{equation}
The matrix elements of $D(z)$ are
\begin{equation}\label{fd4a}
\begin{aligned}
d_{11}(z)&=z^3+\eta_1\eta_2 z^2-
\left(\eta_1^2+\eta_2^2-1+\frac{2\eta_1\eta_2}{\tau}\right)z
-\eta_1\eta_2\,,\\
d_{12}(z)&=\frac{i\tau z^3}{2}
+i\left(-1+\frac{\tau}{2\eta_1\eta_2}\right)z^2-\frac{i\tau z}{2}-\frac{i\tau\eta_1\eta_2}{2}\,,\\
d_{21}(z)&=-\frac{2i\eta_1\eta_2 z}{\tau}+2i\eta_1\eta_2,\quad
d_{22}(z)=z^2-\tau z.
\end{aligned}
\end{equation}
and
\begin{equation}\label{fd4b}
\begin{aligned}
\det D(z)=(z-\tau)(z^2-\eta_1^2)(z^2-\eta_2^2).
\end{aligned}
\end{equation}

{\it Complex conjugation of $\al(z)$ and $\be(z)$.} Formulae \eqref{fd4a} imply that
\begin{equation}\label{sab1}
D(\overline{ z})=\sg_3\overline{D(z)}\sg_3,
\end{equation}
hence by \eqref{fd4}, \eqref{th2},
\begin{equation}\label{sab2}
\phi_{\pm}(\overline{ z})=\sg_3\overline{\phi_{\pm}(z)}\sg_3,
\end{equation}
and by \eqref{th10},
\begin{equation}\label{sab3}
\theta_{\pm}(\overline{ z})=\sg_3\overline{\phi_{\pm}(z)}\sg_3,
\end{equation}
and by \eqref{fdet3},
\begin{equation}\label{sab4}
\al(\overline{ z})=\sg_3\overline{\al(z)}\sg_3,\quad \be(\overline{ z})=\sg_3\overline{\be(z)}\sg_3.
\end{equation}

The following theorem gives the asymptotics of the coefficients $\al_k$, $\be_k$:

\begin{theo} \label{abk} Assume that $0<t<\frac{1}{2}\,$. Then as $k\to\infty$, $\al_k$, $\be_k$ admit
the asymptotic expansions
\begin{equation}\label{fd5}
\al_k\sim \frac{e^{-k\ln\eta_2}}{\sqrt k}\sum_{j=0}^\infty \frac{a_j^0+(-1)^k a_j^1}{k^j}\,,\quad
\be_k\sim \frac{e^{-k\ln\eta_2}}{\sqrt k}\sum_{j=0}^\infty \frac{b_j^0+(-1)^k b_j^1}{k^j}\,.
\end{equation}
The leading coefficients, with $j=0$, are given explicitly as
\begin{equation}\label{fd6}
\begin{aligned}
a_0^0&=\gamma_1 \left[\adj D(\eta_2^{-1})\right]\,\sg_3
\,\left[\adj D^{\rm T}(\eta_2)\right],\\
a_0^1&=\gamma_1[\adj D(-\eta_2^{-1})]\,\sg_3
\,[\adj D^{\rm T}(-\eta_2)],\\
b_0^0&=\gamma_2
\left[D(\eta_2^{-1})\,\sg_3
\, D^{\rm T}(\eta_2)\right],\\
b_0^1&=\gamma_2
\left[ D(-\eta_2^{-1})\,\sg_3
\, D^{\rm T}(-\eta_2)\right],
\end{aligned}
\end{equation}
where
\begin{equation}\label{fd7}
\begin{aligned}
\gamma_1&=
\frac{\gamma_0}{8\sqrt{2\pi}\,\eta_1\eta_2\tau}\,,\quad
\gamma_2=
\frac{\sqrt{2}\,\eta_1\eta_2\tau \gamma_0}{\sqrt{\pi}}\,,\\
\gamma_0&=\frac{1}{(\tau\eta_2-1)
\sqrt{(\eta_1^2-\eta_2^{-2})(\eta_2^2-\eta_2^{-2})(\eta_1^2-\eta_2^2)}}\,.
\end{aligned}
\end{equation}
\end{theo}
A proof of Theorem \ref{abk} is given in Appendix \ref{Appendix_C} below. As a corollary of 
asymptotic formulae \eqref{fd5}, we have the estimates,
\begin{equation}\label{fd8}
|\al_k|\le  \gamma_3 e^{-k\ln\eta_2}\,,\quad
|\be_k|\le \gamma_3 e^{-k\ln\eta_2}\,;\quad \gamma_3>0.
\end{equation}
Substituting these estimates into \eqref{fd1}, we obtain that
\begin{equation}\label{fd9}
\begin{aligned}
|\La_{jk}|\le 
\gamma_4 e^{-(2n+j+k)\ln\eta_2}\,, \quad \gamma_4>0.
\end{aligned}
\end{equation}
Hence
\begin{equation}\label{fd10}
\begin{aligned}
\|\La\|_1\equiv \sum_{j,k=0}^\infty |\La_{jk}|\le 
\gamma_5 e^{-2n\ln\eta_2}\,, \quad \gamma_5>0.
\end{aligned}
\end{equation}
This implies that
\begin{equation}\label{fd11}
\ln\det(1- \La)=-\sum_{k=1}^\infty \frac{\Tr \La^k}{k}=-\Tr\La+\mathcal O(e^{-4n\ln\eta_2}).
\end{equation}
Let us evaluate $\Tr\La$. By \eqref{fd1},
\begin{equation}\label{fd12}
\begin{aligned}
\Tr\La=\sum_{k,a=0}^\infty \Tr (\al_{k+n+a+1}\be_{k+n+a+1})=\sum_{k=1}^\infty k\Tr (\al_{k+n}\be_{k+n}).
\end{aligned}
\end{equation}
To simplify notations, denote
\begin{equation}\label{fd12a}
a_j=a_j^0+(-1)^{k+n} a_j^1,\quad  b_j=b_j^0+(-1)^{k+n} b_j^1.
\end{equation}
Then from \eqref{fd5} we have that for $k\ll n$,
\begin{equation}\label{fd13}
\begin{aligned}
\al_{k+n}&\sim
\frac{e^{-(k+n)\ln\eta_2}}{\sqrt {k+n}}\sum_{j=0}^\infty \frac{a_j}{(k+n)^j}\\
&\sim \frac{e^{-(k+n)\ln\eta_2}}{\sqrt {n}}\left (1-\frac{k}{2n}+\ldots\right)
\left[a_0+\sum_{j=1}^\infty \frac{a_j}{n^j}\left(1-\frac{jk}{n}
+\ldots\right)\right]\\
&\sim \frac{e^{-(k+n)\ln\eta_2}}{\sqrt {n}}\left (a_0+\frac{a_1
-\frac{a_0 k}{2}}{n}+\ldots\right),
\end{aligned}
\end{equation}
and similarly,
\begin{equation}\label{fd13a}
\begin{aligned}
\be_{k+n}&\sim \frac{e^{-(k+n)\ln\eta_2}}{\sqrt {n}}\left (b_0+\frac{b_1
-\frac{b_0 k}{2}}{n}+\ldots\right).
\end{aligned}
\end{equation}
Substituting these formulae into \eqref{fd12}, we obtain that as $n\to\infty$,
\begin{equation}\label{fd14}
\begin{aligned}
\Tr\La\sim \frac{e^{-2n\ln\eta_2}}{n}\sum_{j=0}^\infty \frac{c_j}{n^j}\,,
\end{aligned}
\end{equation}
with
\begin{equation}\label{fd15}
\begin{aligned}
c_0&=\sum_{k=1}^\infty k e^{-2k\ln\eta_2}\Tr (a_0b_0)
=\sum_{k=1}^\infty k e^{-2k\ln\eta_2}\Tr [(a_0^0+(-1)^{k+n} a_0^1)(b_0^0+(-1)^{k+n} b_0^1)]\\
&=\Tr (a_0^0b_0^0+ a_0^1 b_0^1)\sum_{k=1}^\infty k e^{-2k\ln\eta_2}
+(-1)^n\Tr (a_0^0b_0^1+a_0^1b_0^0)\sum_{k=1}^\infty (-1)^k k e^{-2k\ln\eta_2}\,.
\end{aligned}
\end{equation}
We have the identities,
\begin{equation}\label{fd15a}
\begin{aligned}
\sum_{k=1}^\infty k e^{-kt}=\frac{e^{t}}{(e^{t}-1)^2}\,,\quad
\sum_{k=1}^\infty (-1)^k k e^{-kt}=-\frac{e^{t}}{(e^{t}+1)^2}\,,
\end{aligned}
\end{equation}
hence
\begin{equation}\label{fd15b}
\begin{aligned}
c_0&=C_1+(-1)^{n+1}C_2
\end{aligned}
\end{equation}
with
\begin{equation}\label{fd15c}
\begin{aligned}
C_1=\frac{\eta_2^2\Tr (a_0^0b_0^0+ a_0^1 b_0^1)}{(\eta_2^2-1)^2}\,,\quad
C_2=\frac{\eta_2^2\Tr (a_0^0b_0^1+a_0^1b_0^0)}{(\eta_2^2+1)^2}\,.
\end{aligned}
\end{equation}
From \eqref{fd14} we obtain now that
\begin{equation}\label{fd15d}
\begin{aligned}
\Tr\La=\frac{e^{-2n\ln\eta_2}}{n}\Big( c_0+\mathcal O(n^{-1})\Big)\,,
\end{aligned}
\end{equation}
and returning back to formula \eqref{bo8}, we obtain the following asymptotics of the monomer-monomer correlation function:

\begin{theo} \label{th_1} Let $0<t<\frac{1}{2}\,$. Then as $n\to\infty$,
\begin{equation}\label{fd16}
K_2(x,y)=K_2(\infty)\Bigg[1-\frac{e^{-2n\ln\eta_2}}{2n}\,\Big(C_1+(-1)^{n+1}C_2+\mathcal O(n^{-1})\Big)\Bigg], 
\end{equation}
where $C_1$, $C_2$ are defined in \eqref{fd15c}, \eqref{fd6}.
\end{theo}

This gives that the correlation length is equal to
\begin{equation}\label{fd17}
\xi=\frac{1}{2\ln\eta_2}\,.
\end{equation}
As $t\to 0$, $\xi$ diverges as
\begin{equation}\label{fd18}
\xi=\frac{1}{2t}+\mathcal O(1)\,.
\end{equation}

{\it Asymptotics of the constants $C_1$, $C_2$ as $t\to 0$.} From \eqref{p6}, \eqref{p8} we obtain that as $t\to 0$,
\begin{equation}\label{c1}
\eta_1=\sqrt{2}+1+\mathcal O(t^2),\quad \eta_2=1+t+\mathcal O(t^2).
\end{equation}
\begin{center}
\begin{figure}[h]
\begin{center}
\scalebox{0.25}{\includegraphics{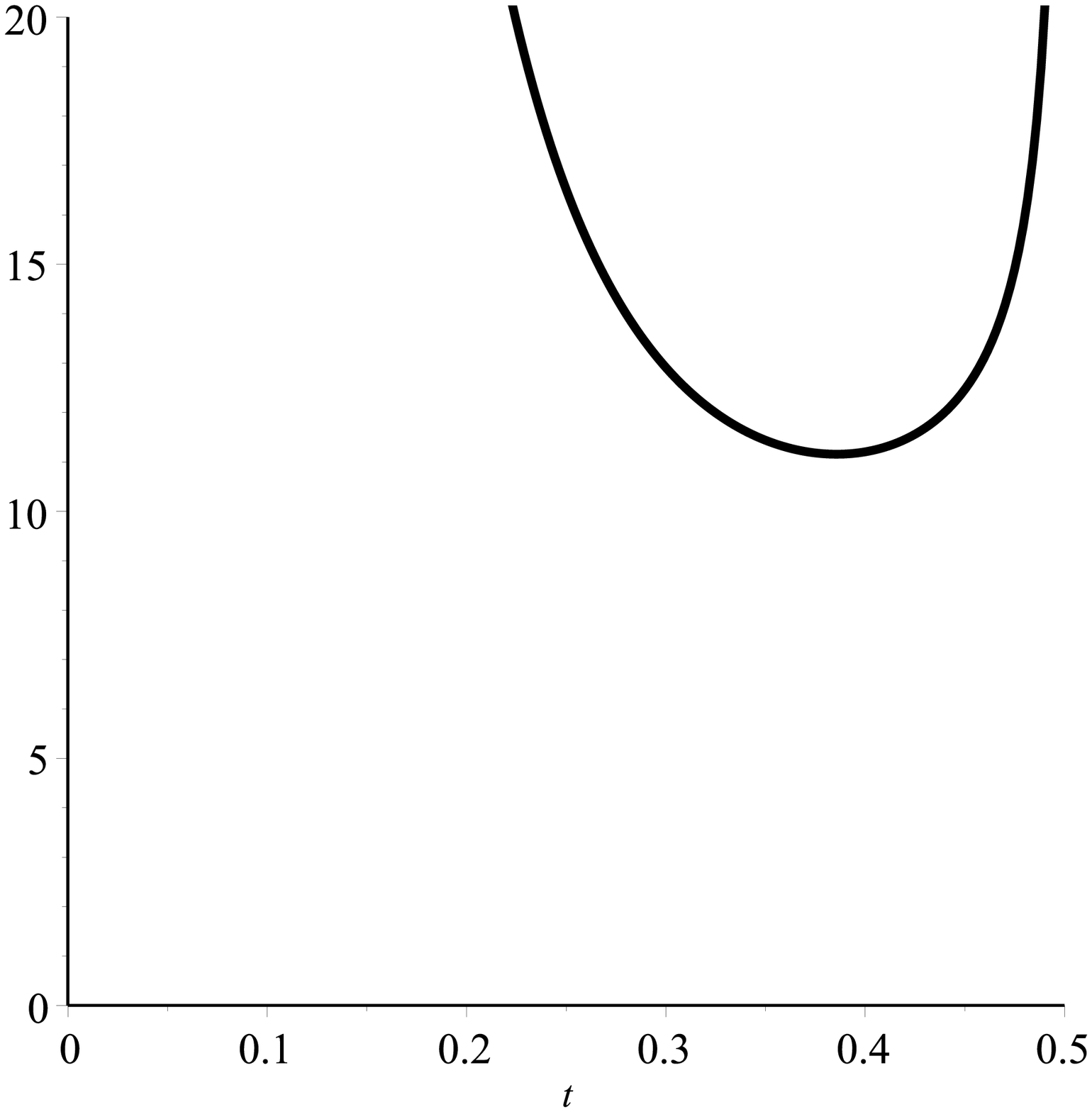}}\hskip 20mm \scalebox{0.25}{\includegraphics{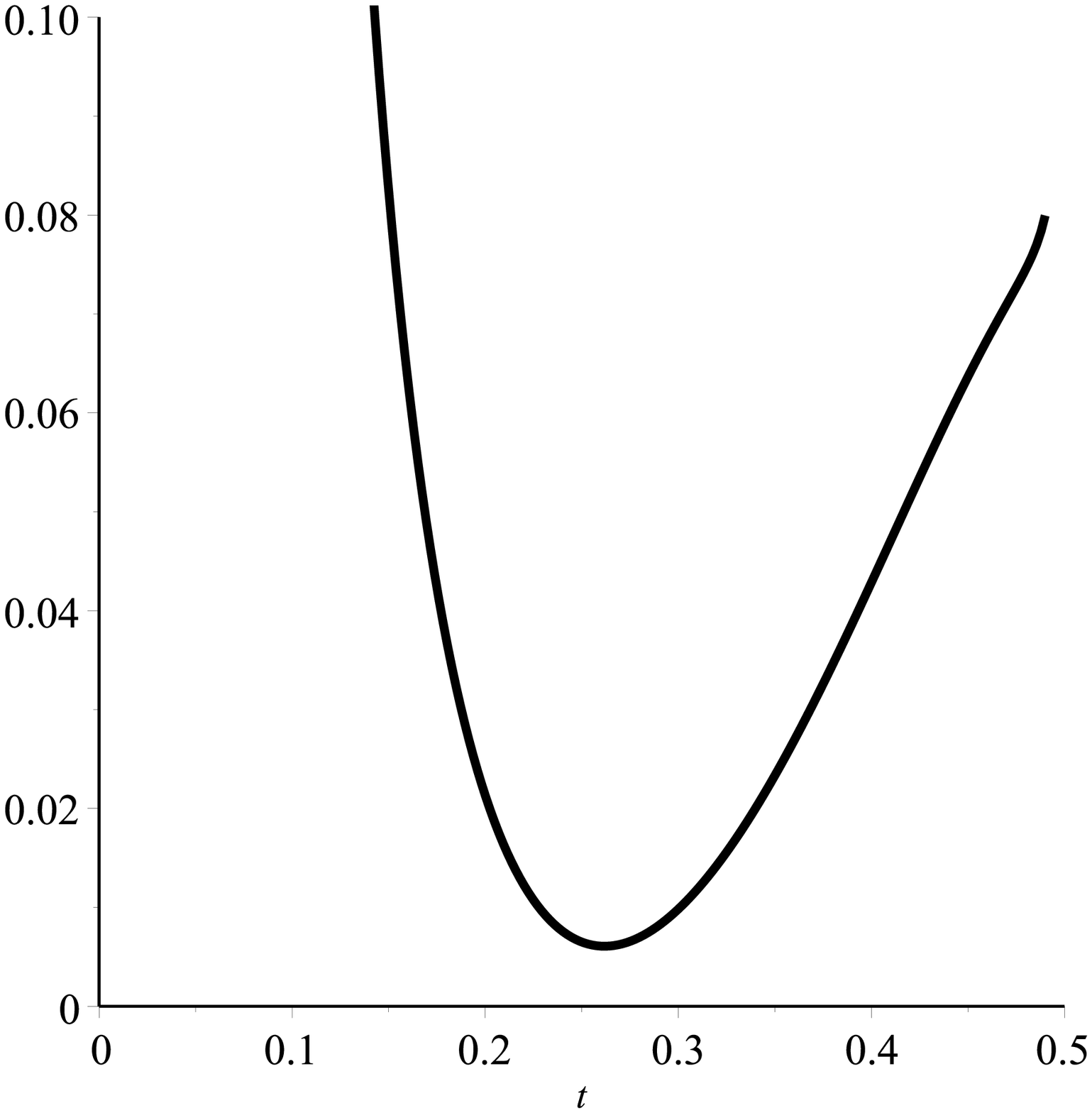}}
\end{center}
  \caption[sectors ]{The graphs of $C_1(t)$, on the left, and $C_2(t)$, on the right.}
 \end{figure}
\label{C(t)}
\end{center}
Substituting these expressions into \eqref{fd15c}, we obtain that as $t\to 0$,
\begin{equation}\label{c9}
C_1=\frac{3-2\sqrt{2}}{128\pi t^5}+\mathcal O(t^{-4}),\quad C_2=\frac{3-2\sqrt{2}}{128\pi t^3}+\mathcal O(t^{-2})\,.
\end{equation}
The graphs of $C_1(t)$, $C_2(t)$ are shown on Fig.3 (observe different scales on the $y$-axis).

\section{Asymptotics of the monomer-monomer correlation function. Supercritical case, $\frac{1}{2} < t < 1$}\label{super_case}

If $t>\frac{1}{2}\,$, then $\eta_1$, $\eta_2$ are complex conjugate numbers,
\begin{equation}\label{scr0}
\eta_{1}=e^{s- i\theta},\quad \eta_{2}= e^{s+ i\theta};\quad s=\ln|\eta_1|=\ln|\eta_2|>0;\quad 0<\theta<\frac{\pi}{4}\,. 
\end{equation} 
The following theorem gives the asymptotics of the coefficients $\al_k$, $\be_k$ in the supercritical case:

\begin{theo} \label{abk_sup} Assume that $\frac{1}{2} < t < 1\,$. Then 
\begin{equation}\label{scr1}
\al_k= \sum_{p=1,2;\,\sg=\pm 1}\al_k(p,\sg),\quad
\be_k= \sum_{p=1,2;\,\sg=\pm 1}\be_k(p,\sg),\quad
\end{equation}
where as $k\to\infty$, $\al_k(p,\sg)$, $\be_k(p,\sg)$ admit
the asymptotic expansions
\begin{equation}\label{scr2}
\al_k(p,\sg)\sim \frac{\sg^{k}
e^{-k\ln \eta_p}}{\sqrt k}\sum_{j=0}^\infty \frac{a_j(p,\sg)}{k^j}\,,\quad
\be_k(p,\sg)\sim \frac{\sg^k e^{-k\ln \eta_p}}{\sqrt k}\sum_{j=0}^\infty \frac{b_j(p,\sg)}{k^j}\,.
\end{equation}
The coefficients $a_j(p,\sg)$, $b_j(p,\sg)$ satisfy the complex conjugacy conditions,
\begin{equation}\label{scr3}
\overline{a_j(2,\sg)}=\sg_3 a_j(1,\sg)\sg_3\,,\quad
\overline{b_j(2,\sg)}=\sg_3 b_j(1,\sg)\sg_3\,.
\end{equation}
The leading coefficients, with $j=0$, are equal to
\begin{equation}\label{scr4}
\begin{aligned}
&a_0(p,\sg)=\gamma_p\,[\adj D(\sg\eta_p^{-1})]\,\sg_3\,[\adj D^{\rm T}(\sg\eta_p)],\\
&b_0(p,\sg)=\ep_p\, [D(\sg\eta_p^{-1})]\,\sg_3\, [D^{\rm T}(\sg\eta_p)],
\end{aligned}
\end{equation}
where
\begin{equation}\label{scr5}
\begin{aligned}
&\gamma_p=\frac{1}{8\sqrt{2\pi}\,\eta_1\eta_2\tau
(\tau \eta_p-1)\sqrt{(\eta_1^2-\eta_p^{-2})(\eta_2^2-\eta_p^{-2})(\eta_{p'}^2-\eta_p^2)}}\,,\\
&\ep_p=\frac{\sqrt{2}\,\eta_1\eta_2\tau }{\sqrt{\pi}\,(\tau \eta_p-1)
\sqrt{(\eta_1^2-\eta_p^{-2})(\eta_2^2-\eta_p^{-2})(\eta_{p'}^2-\eta_p^2)}},
\quad p'=2-p.
\end{aligned}
\end{equation}
\end{theo}

A proof of Theorem \ref{abk_sup} is given below in Appendix \ref{Appendix_D}.
As a corollary of 
asymptotic formulae \eqref{scr2}, we have the exponential estimates,
\begin{equation}\label{scr6}
|\al_k|\le  C_0 e^{-k s}\,,\quad
|\be_k|\le C_0 e^{-k s}\,;\quad s=\ln|\eta_1|=\ln|\eta_2|,\quad C_0>0, 
\end{equation}
which imply that
\begin{equation}\label{scr7}
\ln\det(1- \La)=-\Tr\La+\mathcal O(e^{-4s n}),
\end{equation}
cf. \eqref{fd11}. By \eqref{fd12},
\begin{equation}\label{scr8}
\begin{aligned}
\Tr\La=\sum_{k=1}^\infty k\Tr (\al_{k+n}\be_{k+n}).
\end{aligned}
\end{equation}
To simplify notations, denote
\begin{equation}\label{scr9}
a_j=\sum_{p=1,2;\,\sg=\pm 1} \sg^k e^{i(-1)^p\theta k} a_j(p,\sg),\quad
b_j=\sum_{p=1,2;\,\sg=\pm 1}\sg^k e^{i(-1)^p\theta k} b_j(p,\sg).
\end{equation}
Then \eqref{scr1}, \eqref{scr2} give that
\begin{equation}\label{scr10}
\al_k\sim \frac{
e^{-k s}}{\sqrt k}\sum_{j=0}^\infty \frac{a_j}{k^j}\,,\quad
\be_k\sim \frac{e^{-k s}}{\sqrt k}\sum_{j=0}^\infty \frac{b_j}{k^j}\,,
\end{equation}
hence for $k\ll n$,
\begin{equation}\label{scr11}
\begin{aligned}
\al_{k+n}&\sim
\frac{e^{-(k+n)s}}{\sqrt {k+n}}\sum_{j=0}^\infty \frac{a_j}{(k+n)^j}\\
&\sim \frac{e^{-(k+n)s}}{\sqrt {n}}\left (1-\frac{k}{2n}+\ldots\right)
\left[a_0+\sum_{j=1}^\infty \frac{a_j}{n^j}\left(1-\frac{jk}{n}
+\ldots\right)\right]\\
&\sim \frac{e^{-(k+n)s}}{\sqrt {n}}\left (a_0+\frac{a_1
-\frac{a_0 k}{2}}{n}+\ldots\right),
\end{aligned}
\end{equation}
Similarly,
\begin{equation}\label{scr12}
\begin{aligned}
\be_{k+n}&\sim \frac{e^{-(k+n)s}}{\sqrt {n}}\left (b_0+\frac{b_1
-\frac{b_0 k}{2}}{n}+\ldots\right).
\end{aligned}
\end{equation}
Substituting these formulae in \eqref{scr8}, we obtain that as $n\to \infty$,
\begin{equation}\label{scr13}
\begin{aligned}
\Tr\La&=\sum_{k=1}^\infty k\Tr (\al_{k+n}\be_{k+n})
=\sum_{k=1}^\infty k\Tr \left[\frac{e^{-(k+n)s}}{\sqrt {n}}\left (a_0+\frac{a_1
-\frac{a_0 k}{2}}{n}+\ldots\right)\right.\\
&\times\left. \frac{e^{-(k+n)s}}{\sqrt {n}}\left (b_0+\frac{b_1
-\frac{b_0 k}{2}}{n}+\ldots\right)\right]=\frac{e^{-2ns}}{n}\sum_{j=0}^\infty \frac{c_j}{n^j}\,,
\end{aligned}
\end{equation}
with 
\begin{equation}\label{scr14}
\begin{aligned}
c_0&=\sum_{k=1}^\infty ke^{-2kr}\Tr(a_0b_0).
\end{aligned}
\end{equation}
By \eqref{scr9},
\begin{equation}\label{scr15}
\begin{aligned}
a_0&=\sum_{p=1,2;\,\sg=\pm 1} \sg^{k+n} e^{i(-1)^p\theta (k+n)} a_0(p,\sg)
=e^{-i\theta (k+n)}a_0(1,1)+e^{i\theta (k+n)}a_0(2,1)\\
&+(-1)^{k+n} e^{-i\theta (k+n)}a_0(1,-1)
+(-1)^{k+n} e^{i\theta (k+n)}a_0(2,-1),
\end{aligned}
\end{equation}
and similarly,
\begin{equation}\label{scr16}
\begin{aligned}
b_0&=e^{-i\theta (k+n)}b_0(1,1)+e^{i\theta (k+n)}b_0(2,1)\\
&+(-1)^{k+n} e^{-i\theta (k+n)}b_0(1,-1)
+(-1)^{k+n} e^{i\theta (k+n)}b_0(2,-1)\\
\end{aligned}
\end{equation}
hence
\begin{equation}\label{scr17}
\begin{aligned}
a_0b_0&=e^{-2i\theta (k+n)}\Big(c_{11}+(-1)^{k+n}c_{12}\Big)
+\Big(c_{21}+(-1)^{k+n}c_{22}\Big)\\
&+e^{2i\theta (k+n)}\Big(c_{31}+(-1)^{k+n}c_{32}\Big),
\end{aligned}
\end{equation}
with
\begin{equation}\label{scr18}
\begin{aligned}
c_{11}&=a_0(1,1)b_0(1,1)+a_0(1,-1)b_0(1,-1),\\ c_{12}&=a_0(1,1)b_0(1,-1)+a_0(1,-1)b_0(1,1),\\
c_{21}&=a_0(1,1)b_0(2,1)+a_0(2,1)b_0(1,1)+a_0(1,-1)b_0(2,-1)+a_0(2,-1)b_0(1,-1),\\
c_{22}&=a_0(1,1)b_0(2,-1)+a_0(2,-1)b_0(1,1)+a_0(2,1)b_0(1,-1)+a_0(1,-1)b_0(2,1),\\
c_{31}&=a_0(2,1)b_0(2,1)+a_0(2,-1)b_0(2,-1),\\ c_{32}&=a_0(2,1)b_0(2,-1)+a_0(2,-1)b_0(2,1).
\end{aligned}
\end{equation}
From \eqref{scr3} we obtain the complex conjugacy conditions as
\begin{equation}\label{scr18a}
\begin{aligned}
\overline{c_{11}}=\sg_3 c_{31}\sg_3,\quad 
\overline{c_{12}}=\sg_3 c_{32}\sg_3,\quad 
\overline{c_{21}}=\sg_3 c_{21}\sg_3,\quad 
\overline{c_{22}}=\sg_3 c_{22}\sg_3. 
\end{aligned}
\end{equation}
Using the formula
\begin{equation}\label{scr19}
\sum_{k=1}^\infty ke^{-\la k}=\frac{e^\la}{(e^\la-1)^2}\,,\quad \Re\la>0,
\end{equation}
we obtain that
\begin{equation}\label{scr20}
\begin{aligned}
c_0&=\sum_{k=1}^\infty ke^{-2ks}\Tr(a_0b_0)
=\sum_{k=1}^\infty ke^{-2ks}\Tr\Big[e^{-2i\theta (k+n)}\Big(c_{11}+(-1)^{k+n}c_{12}\Big)\\
&+\Big(c_{21}+(-1)^{k+n}c_{22}\Big)
+e^{2i\theta (k+n)}\Big(c_{31}+(-1)^{k+n}c_{32}\Big)\Big]\\
&=e^{-2i\theta n}d_{11}+(-1)^n e^{-2i\theta n}d_{12}+d_{21}+(-1)^n d_{22}+e^{2i\theta n}d_{31}+
(-1)^n e^{2i\theta n}d_{32}\,,
\end{aligned}
\end{equation}
with
\begin{equation}\label{scr21}
\begin{aligned}
d_{11}&=\frac{e^{2s+2i\theta}\Tr c_{11}}{(e^{2s+2i\theta}-1)^2}\,,\quad 
d_{12}=-\frac{e^{2s+2i\theta}\Tr c_{12}}{(e^{2s+2i\theta}+1)^2}\,,\\ 
d_{21}&=\frac{e^{2s}\Tr c_{21}}{(e^{2s}-1)^2}\,,\quad 
d_{22}=-\frac{e^{2s}\Tr c_{22}}{(e^{2s}+1)^2}\,,\\ 
d_{31}&=\frac{e^{2s-2i\theta}\Tr c_{31}}{(e^{2s-2i\theta}-1)^2}\,,\quad 
d_{32}=-\frac{e^{2s-2i\theta}\Tr c_{32}}{(e^{2s-2i\theta}+1)^2}\,.
\end{aligned}
\end{equation}
By \eqref{scr18a},
\begin{equation}\label{scr22}
\begin{aligned}
\overline{d_{11}}=d_{31},\quad 
\overline{d_{12}}=d_{32},\quad 
\overline{d_{21}}=d_{21},\quad 
\overline{d_{22}}=d_{22}, 
\end{aligned}
\end{equation}
hence
\begin{equation}\label{scr23}
\begin{aligned}
&e^{-2i\theta n}d_{11}+e^{2i\theta n}d_{31}=C_1\cos(2\theta n+\varphi_1),\\
&(-1)^n e^{-2i\theta n}d_{12}+(-1)^n e^{2i\theta n} d_{32}=C_2(-1)^n\cos(2\theta n+\varphi_2),
\end{aligned}
\end{equation}
with 
\begin{equation}\label{scr23a}
\begin{aligned}
C_1=2|d_{11}|,\quad \f_1=-\arg d_{11};\quad C_2=2|d_{12}|,\quad \f_1=-\arg d_{12}.
\end{aligned}
\end{equation}
Thus, by \eqref{scr20},
\begin{equation}\label{scr24}
\begin{aligned}
c_0=C_1\cos(2\theta n+\varphi_1)+C_2(-1)^n\cos(2\theta n+\varphi_2)+C_3+C_4(-1)^n,
\end{aligned}
\end{equation}
where
\begin{equation}\label{scr25}
\begin{aligned}
C_3=d_{21},\quad C_4=d_{22}.
\end{aligned}
\end{equation}
Returning back to formula \eqref{scr13}, we obtain that
\begin{equation}\label{scr26}
\begin{aligned}
\Tr\La&=\frac{e^{-2ns}}{n}\,\Big(C_1\cos(2\theta n+\varphi_1)+C_2(-1)^n\cos(2\theta n+\varphi_2)+C_3+C_4(-1)^n\\
&+\mathcal O(n^{-1})\Big).
\end{aligned}
\end{equation}
This implies the following asymptotic formula for $K_2(x,y)$:

\begin{theo}\label{th_2}
Let $\frac{1}{2} < t < 1\,$. Then as $n\to\infty$,
\begin{equation}\label{scr27}
\begin{aligned}
K_2(x,y)&=K_2(\infty)\Bigg[1-\frac{e^{-2ns}}{2n}\,\Big(C_1\cos(2\theta n+\varphi_1)+C_2(-1)^n\cos(2\theta n+\varphi_2)\\
&+C_3+C_4(-1)^n\Big)+\mathcal O(n^{-1})\Bigg], 
\end{aligned}
\end{equation}
with $s=\ln|\eta_1|=\ln|\eta_2|$, $\theta=|\arg\eta_1|=|\arg\eta_2|$, and $C_1$, $C_2$, $C_3$ , $C_4$, $\f_1$, $\f_2$ given
in \eqref{scr23a}, \eqref{scr25}.
\end{theo}

{\it The limiting value of the constants as $t\to 1$.} When $t=1$ the dimer model has a symmetry
of the dihedral group $D_6$. As $t\to 1$, we find the
limiting values of the constants in \eqref{scr27} to be equal to
\begin{equation}\label{t11}
\begin{aligned}
&K_2(\infty)=\frac{1}{6}\sqrt{6-3\sqrt{3}}=0.149429\ldots,\\
&\xi:= \frac{1}{2s}=0.601344\ldots,\quad
 \omega := 2\theta=0.749469\ldots, \\
&\frac{C_1}{2}=11.769354\ldots,\quad \frac{C_2}{2}=2.014601\ldots,\\
&\frac{C_3}{2}=30.674027\ldots,\quad \frac{C_4}{2}=0.676743\ldots,\\
&\f_1=-1.556067\ldots,\quad \f_2=-2.282124\ldots .
\end{aligned}
\end{equation}
Observe that when $t=1$, the symbol $\phi(z)$ in \eqref{mm5} becomes singular. This indicates that
$t=1$ is a critical point of the model and some additional subdominant terms in the asymptotics of $K_2(x,y)$
should appear. A calculation of these terms is an interesting, challenging
problem, and we leave this problem for the future research.

\appendix

\section{The Borodin--Okounkov--Case--Geronimo type formula for block Toeplitz determinants}\label{Appendix_A}

We begin with some preliminary facts about Toeplitz operators and Toeplitz matrices. Let $\phi \in L^{\infty}(\mathbb{T})^{N \times N}$ be an essentially bounded $N \times N$ matrix valued function defined on the unit circle with Fourier coefficients $\phi_k\in \C^{N\times N}$.
The Toeplitz and Hankel operators are defined on $\ell^{2}(\Z_+)$, $\Z_+=\{0,1,\dots\}$, by means of the semi-infinite infinite matrices
\begin{eqnarray*}
 T(\phi) & =&  (\phi_{j-k}), \,\,\,\,\,\,\,\,\,\,\,\,\,0 \leq j,k < \infty,\\[1ex]
 H(\phi) &=&  (\phi_{j+k+1}), \,\,\,\,\,\,\,\,0 \leq j,k < \infty.
 \end{eqnarray*}
For $\phi, \psi \in L^{\infty}(\mathbb{T})^{N \times N}$ the identities
\bq\label{fT1}
T(\phi \psi) & =& T(\phi)T(\psi) +H(\phi)H({\tilde \psi})\\
H(\phi \psi) & =& T(\phi)H(\psi) +H(\phi)T({\tilde \psi})
\eq
are well-known. It follows from these identities that if  $\psi_{-}$ and $\psi_{+} $ have the property that all their Fourier coefficients vanish for $k > 0$ and $k < 0$, respectively, then 
\bq\label{fT2}
T(\psi_{-} \phi \psi_{+} )& =& T(\psi_{-} )T(\phi) T( \psi_{+} ) ,\\
H(\psi_{-} \phi \tilde{\psi}_{+} )& =& T(\psi_{-} )H(\phi) T( \psi_{+} ) .
\label{THT}
\eq

We 
let $\B$ stand for the set of all function $\phi\in L^1(\T)$ such that the Fourier coefficients satisfy
 \begin{equation}\label{B}
 \| \phi\|_{\B} :=\sum_{k =-\infty} ^{\infty} |\phi_{k}| + 
 \Big(\sum_{k = -\infty}^{\infty} |k|\cdot|\phi_{k}|^{2}\Big)^{1/2} < \infty.
 \end{equation}
With the norm (\ref{B}) and pointswise defined algebraic operations on $\T$, the set $\B$ becomes a Banach algebra of continuous functions on the unit circle. This is a convenient algebra because, among other things, the Hankel operators with symbols from this algebra are Hilbert-Schmidt operators and thus the product of two of them is always trace class.

In what follows all determinants are defined on the images of the appropriate projections.

Suppose that $\psi\in\B^{N\times N}$ such that both $T(\psi)$ and $T(\tilde{\psi})$ are invertible on $(\ell^2(\Z_+))^N$.
Then
\bq\label{f2}
\det T_n(\psi^{-1}) &=& \frac{E(\psi)}{G(\psi)^n}\cdot 
\det\left(I-H(z^{-n}v_{-}u_{+}^{-1})H(\tilde{u}_{-}^{-1} \tilde{v}_{+}z^{-n})\right),
\eq
where $$
\psi^{-1}=u_{-}u_{+}=v_{+}v_{-}.
$$

\begin{proof}
The invertibility of $T(\psi)$ and $T(\tilde{\psi})$ implies the invertibility of $\psi$ and  the existence of a left and a right  canonical factorization (in $\B^{N \times N}$)
$$
\psi^{-1}=u_{-}u_{+}=v_{+}v_{-}.
$$
The first step is to verify that 
$$
\det T_{n}(\psi^{-1}) =G(\psi^{-1})^{n}\cdot\det P_{n}T(v_{+}^{-1}) T(u_{-})T(u_{+})T(v_{-}^{-1})P_{n},
$$
where $P_n$ is the projection $\{x_k\}_{k=0}^\infty\mapsto \{x_0,\dots,x_{n-1},0,0, \dots\}$ acting on 
$\ell^2(\Z_+)$. 
This is because
$$
\det T_{n}(\psi^{-1}) = \det P_{n}T(u_{-}u_{+}) P_{n} = \det P_{n}T(v_{+}) T(v_{+}^{-1}) P_{n} T(u_{-})T(u_{+}) P_{n} T(v_{-}^{-1}) T(v_{-}) P_{n}$$
or
$$
\det T_{n}(\psi^{-1}) = \det P_{n}T(u_{-}u_{+}) P_{n} = \det P_{n}T(v_{+})P_{n} T(v_{+}^{-1})  T(u_{-})T(u_{+}) T(v_{-}^{-1}) P_{n}T(v_{-}) P_{n}.$$
This last formula follows since the matrices $T(v_{+})$ and $T(v_{-})$ are lower and upper triangular block matrices.

Then we notice that $\det P_{n}T(v_{+})P_{n} \times \det P_{n}T(v_{-}) P_{n}  = G(\psi^{-1})^{n} $ and this step is verified.
To obtain the next step we use the Jacobi identity 
\begin{eqnarray}
\det P_{n}A P_{n} &=& \det (Q_{n}+P_{n}A P_{n})=\det (Q_{n}+A P_{n})\nn\\
&=& \det A\cdot \det (A^{-1} Q_{n}+P_{n})=
\det A\cdot\det (Q_{n}A^{-1} Q_{n})\label{f.K}
\end{eqnarray}
in which $Q_{n}=I-P_{n}$ and $A$ is an invertible operator of the form identity plus trace class.

 We apply this to $A =  T(v_{+}^{-1}) T(u_{-})T(u_{+})T(v_{-}^{-1}) = T(v_{+}^{-1}) T(\psi^{-1})T(v_{-}^{-1}) $. It  follows that
\begin{eqnarray*}
\lefteqn{\det P_{n}T(v_{+}^{-1}) T(u_{-})T(u_{+})T(v_{-}^{-1})P_{n}}
\hspace{6ex}\\[1ex]
&=&
\det T(v_{+}^{-1}) T(u_{-})T(u_{+})T(v_{-}^{-1})\cdot
\det Q_{n}T(v_{-})T(u_{+}^{-1})T(u_{-}^{-1})T(v_{+})Q_{n}.
\end{eqnarray*}
The first term on the right is easily seen to be $\det T(\psi)T(\psi^{-1}) = E(\psi).$ Notice that by the factorization 
$$v_{-} u_{+}^{-1} u_{-}^{-1} v_{+} =  I_{N}$$ and thus
$$\det Q_{n}T(v_{-})T(u_{+}^{-1})T(u_{-}^{-1})T(v_{+})Q_{n} = \det (P_{n }+Q_{n}T(v_{-} u_{+}^{-1})T(u_{-}^{-1} v_{+})Q_{n}) $$
$$=  \det( P_{n} + Q_{n}(I - H(v_{-}u_{+}^{-1})H(\tilde{u}_{-}^{-1} \tilde{v}_{+}))Q_{n} )$$
$$ = \det (I  - T(z^{n})T(z^{-n}) H(v_{-}u_{+}^{-1})H(\tilde{u}_{-}^{-1} \tilde{v}_{+})T(z^{n})T(z^{-n}))$$
using the fact that $Q_{n} = T(z^{n})T(z^{-n}).$
(These last computations rely on the formulas given at the beginning of this section.)
Now since 
$$\det ( I + AB ) = \det (I + BA) $$
this last determinant is the same as 
$$ \det( I  - T(z^{-n}) H(v_{-}u_{+}^{-1})H(\tilde{u}_{-}^{-1} \tilde{v}_{+})T(z^{n}))$$
or
$$ \det ( I  - H(z^{-n}v_{-}u_{+}^{-1})H(\tilde{u}_{-}^{-1} \tilde{v}_{+}z^{-n})).$$

To compute the trace of a product of Hankels we simply need to consider the following.
The $j,k$ (block) coefficient of the product of 
$H(A_{1})H(A_{2})$ is given by
$$\sum_{l=0}^{\infty} (A_{1})_{1+j+l} (A_{2})_{1+l+k} $$ and thus the trace is given by
$$ \sum_{l=0}^{\infty}\sum_{k=0}^{\infty} \Tr (A_{1})_{1+k+l} (A_{2})_{1+k+l}. $$

This last expression is the same as 
$$ \sum_{k=1}^{\infty}k \,\Tr (A_{1})_{k} (A_{2})_{k}. $$

Thus we have that 

$$ \Tr H(z^{-n}v_{-}u_{+}^{-1})H(\tilde{u}_{-}^{-1} \tilde{v}_{+}z^{-n})) = \sum_{k=1}^{\infty}k \,\Tr (v_{-}u_{+}^{-1})_{k + n} (\tilde{u}_{-}^{-1} \tilde{v}_{+})_{k +n},$$

since the effect of multiplying by $z^{-n}$ shifts the Fourier coefficients.

\end{proof}

\section{Proof of Theorem \ref{factor}} \label{Appendix_B}

Let us recall that by \eqref{p2}, $\phi(z)=\sg(z)\pi(z)$, where $\sg(z)$ is a scalar function. The difficult part is to factor $\pi(z)$.
To factor $\pi(z)$ we use a {\it decreasing power algorithm}. In this algorithm at every step we
make a substitution decreasing the power in $z$ of the matrix entries under consideration.
 
As the first step, we write $\pi(z)$ as
\begin{equation}\label{app1}
 \pi(z)=\tau^{-2}
\begin{pmatrix}
1 & 0 \\
0 & z-\tau
\end{pmatrix}
\rho(z)
\begin{pmatrix}
z^{-1}-\tau & 0 \\
0 & 1
\end{pmatrix}.
\end{equation}
where
\begin{equation}\label{app2}
 \rho(z)=
\begin{pmatrix}
 \rho_{11}(z) &  \rho_{12}(z) \\
 \rho_{21}(z) &  \rho_{22}(z)
\end{pmatrix}
\end{equation}
with
\begin{equation}\label{app3}
\begin{aligned}
\rho_{11}(z)&=z(\cos x+\tau\sin^2x)
=-\frac{\tau z^3}{4}+\frac{z^2}{2}+\frac{\tau z}{2}+\frac{1}{2}-\frac{\tau }{4z}\,,\\
\rho_{12}(z)&=\sin x\,(z-\tau)(z^{-1}-\tau)
=\frac{i\tau z^2}{2}-\frac{i(\tau^2+1)z}{2}+\frac{i(\tau^2+1)}{2z}-\frac{i\tau}{2z^2}\,,\\
\rho_{21}(z)&=-\sin x=\frac{i(z^2-1)}{2z}=\frac{iz}{2}-\frac{i}{2z}\,,\\
\rho_{22}(z)&=z^{-1}(\cos x+\tau\sin^2x)=
-\frac{\tau z}{4}+\frac{1}{2}+\frac{\tau }{2z}+\frac{1}{2 z^2}-\frac{\tau }{4 z^3}\,
\end{aligned}
\end{equation}
and
\begin{equation}\label{app4}
 \det \rho(z)= \frac{\tau^2}{16 \eta_1^2\eta_2^2}
\left(z^{-2}-\eta_1^2\right)\left( z^{-2}-\eta_2^2\right)
\left(z^2-\eta_1^2\right)\left(z^2-\eta_2^2\right).
\end{equation}

Let us factor $\rho(z)$. Let $p_1$ be a constant.
We have that 
\begin{equation}\label{fac6}
\begin{pmatrix}
1 & p_1 \\
0 & 1
\end{pmatrix}^{-1}\rho(z)=
\begin{pmatrix}
\rho_{11}(z)-p_1 \rho_{21}(z) & \rho_{12}(z)-p_1 \rho_{22}(z) \\
\rho_{21}(z) & \rho_{22}(z)
\end{pmatrix}.
\end{equation}
Let us take
\begin{equation}\label{q1}
p_1=\frac{\rho_{11}(\eta_1)}{\rho_{21}(\eta_1)}\,,
\end{equation}
so that
\begin{equation}\label{q2}
\rho_{11}(\eta_1)-p_1 \rho_{21}(\eta_1)=0.
\end{equation}
Observe that since $|\eta_1|>1$, 
\[
\rho_{21}(\eta_1)=-\frac{i(\eta_1^2-1)}{2\eta_1}\not=0.
\]
Equation \eqref{q2} ensures that $\rho_{11}(z)-p_1 \rho_{21}(z)$ is divisible by $(z-\eta_1)$.
Since by \eqref{app4}, $\det \rho(\eta_1)=0$,  we obtain that
\begin{equation}\label{q3}
\rho_{12}(\eta_1)-p_1 \rho_{22}(\eta_1)=\rho_{12}(\eta_1)-\frac{\rho_{11}(\eta_1)\rho_{22}(\eta_1)}{\rho_{21}(\eta_1)} =0,
\end{equation}
hence $\rho_{12}(z)-p_1 \rho_{22}(z)$ is divisible by $(z-\eta_1)$ as well. 

Consider the matrix $\rho^{(1)}(z)$ with the matrix elements
\begin{equation}\label{q4}
\begin{aligned}
\rho_{11}^{(1)}(z)&=\frac{\rho_{11}(z)-p_1 \rho_{21}(z)}{z-\eta_1}\,,
\quad \rho_{12}^{(1)}(z)=\frac{\rho_{12}(z)-p_1 \rho_{22}(z)}{z-\eta_1}\,,\\
\rho_{21}^{(1)}(z)&=\rho_{21}(z),\quad
\qquad \rho_{22}^{(1)}(z)=\rho_{22}(z),
\end{aligned}
\end{equation}
so that
\begin{equation}\label{q5}
\begin{aligned}
\rho(z)=
\begin{pmatrix}
1 & p_1 \\
0 & 1
\end{pmatrix}
\begin{pmatrix}
z-\eta_1 & 0 \\
0 & 1
\end{pmatrix}
\rho^{(1)}(z).
\end{aligned}
\end{equation}
From \eqref{q1} and \eqref{app3} we obtain that
\begin{equation}\label{q6}
\begin{aligned}
p_1=\frac{i\big[\tau(\eta_1^2-1)^2-2\eta_1(\eta_1^2+1)\big]}{2(\eta_1^2-1)}\,,
\end{aligned}
\end{equation}
and $\rho_{11}^{(1)}(z)$, $\rho_{12}^{(1)}(z)$ are rational functions in $z$ with leading terms at infinity
\begin{equation}\label{q7}
\begin{aligned}
\rho_{11}^{(1)}(z)&
=-\frac{\tau z^2}{4}+\mathcal{O}(z),\quad 
\rho_{12}^{(1)}(z)&=\frac{i\tau z}{2}+\mathcal{O}(1)\,.
\end{aligned}
\end{equation}
Observe that the degrees of the functions $\rho_{11}^{(1)}(z)$, $\rho_{12}^{(1)}(z)$ in $z$ decrease by one comparing 
to $\rho_{11}(z)$, $\rho_{12}(z)$ in \eqref{app3}.
From \eqref{app4} and \eqref{q5} we obtain that 
\begin{equation}\label{q8}
\begin{aligned}
\det \rho^{(1)}(z)=\frac{\det \rho(z)}{z-\eta_1}
= \frac{\tau^2}{16 \eta_1^2\eta_2^2}
\left(z^{-2}-\eta_1^2\right)\left( z^{-2}-\eta_2^2\right)
\left(z+\eta_1\right)\left(z^2-\eta_2^2\right)\,.
\end{aligned}
\end{equation}

Applying the same procedure to $\rho^{(1)}(z)$, we obtain that
\begin{equation}\label{q9}
\begin{aligned}
\rho^{(1)}(z)=
\begin{pmatrix}
1 & p_2 \\
0 & 1
\end{pmatrix}
\begin{pmatrix}
z+\eta_1 & 0 \\
0 & 1
\end{pmatrix}
\rho^{(2)}(z),
\end{aligned}
\end{equation}
with
\begin{equation}\label{q10}
\begin{aligned}
p_2=-\frac{i(\eta_1^2+1)}{\eta_1^2-1}\,,
\end{aligned}
\end{equation}
and 
\begin{equation}\label{q11}
\begin{aligned}
\rho_{11}^{(2)}(z)&
=-\frac{\tau z}{4}-\frac{1}{\eta_1^2-1}+\frac{\tau}{4z},\quad 
\rho_{12}^{(2)}(z)&=\frac{i\tau (\eta_1^2-3)}{4(\eta_1^2-1)}+\mathcal{O}(z^{-1})\,.
\end{aligned}
\end{equation}

Applying the same procedure to $\rho^{(2)}(z)$, we obtain that
\begin{equation}\label{q12}
\begin{aligned}
\rho^{(2)}(z)=
\begin{pmatrix}
1 & p_3 \\
0 & 1
\end{pmatrix}
\begin{pmatrix}
z-\eta_2 & 0 \\
0 & 1
\end{pmatrix}
\rho^{(3)}(z),
\end{aligned}
\end{equation}
with
\begin{equation}\label{q13}
\begin{aligned}
p_3=\frac{i\big[\tau(\eta_1^2-1)(\eta_2^2-1)+4\eta_2\big]}{2(\eta_1^2-1)(\eta_2^2-1)}
=\frac{i\tau(\eta_1+1)}{2\eta_1}\,,
\end{aligned}
\end{equation}
(to simplify we use equation \eqref{p13}) and 
\begin{equation}\label{q14}
\begin{aligned}
\rho_{11}^{(3)}(z)
&=\frac{\tau}{4\eta_1}+\frac{\tau}{4\eta_1\eta_2 z}\,,\\
\rho_{12}^{(3)}(z)
&=\frac{i\tau^2(\eta_1+1)}{8\eta_1}+\mathcal{O}(z^{-1}).
\end{aligned}
\end{equation}
Observe that both $\rho_{11}^{(3)}(z)$ and $\rho_{12}^{(3)}(z)$ are of degree 0 at infinity.

We did not change so far the entries $\rho_{21}(z)$ and $\rho_{22}(z)$ in \eqref{app3} and their degree in $z$ is 1.
It is time to decrease their degree. We obtain that
\begin{equation}\label{q15}
\begin{aligned}
\rho^{(3)}(z)=
\begin{pmatrix}
1 & 0 \\
p_4 & 1
\end{pmatrix}
\begin{pmatrix}
1 & 0 \\
0 & z+\eta_2
\end{pmatrix}
\rho^{(4)}(z),
\end{aligned}
\end{equation}
where
\begin{equation}\label{q16}
\begin{aligned}
p_4=-\frac{2i\eta_1\eta_2}{\tau}\,,
\end{aligned}
\end{equation}
and 
\begin{equation}\label{q17}
\begin{aligned}
\rho^{(4)}(z)=
\begin{pmatrix}
\di\frac{\tau}{4\eta_1} &\di \frac{i\tau^2(\eta_1+1)}{8\eta_1}\\
\di \frac{i}{2} & \di -\frac{\tau}{4}
\end{pmatrix}+\mathcal{O}(z^{-1}),\quad z\to\infty.
\end{aligned}
\end{equation}
From \eqref{q8} we have that
\begin{equation}\label{q18}
\begin{aligned}
\det \rho^{(4)}(z)=\frac{\det \rho(z)}{z-\eta_1}
= \frac{\tau^2}{16 \eta_1^2\eta_2^2}
\left(z^{-2}-\eta_1^2\right)\left( z^{-2}-\eta_2^2\right)\,.
\end{aligned}
\end{equation}
Since $\det \rho^{(4)}(z)\not=0$ for $|z|\ge 1$,  $\rho^{(4)}(z)$ is invertible for $|z|\ge 1$.

In summary, 
\begin{equation}\label{q19}
\pi(z)=v_+(z)v_-(z),
\end{equation}
where
\begin{equation}\label{q20}
\begin{aligned}
v_+(z)=D_0(z)P_1D_1(z)P_2D_2(z)P_3D_3(z)P_4D_4(z),
\end{aligned}
\end{equation}
and
\begin{equation}\label{q23}
\begin{aligned}
v_-(z)=\tau^{-2}\rho^{(4)}(z)
\begin{pmatrix}
z^{-1}-\tau & 0 \\
0 & 1
\end{pmatrix}.
\end{aligned}
\end{equation}
Observe that $v_+(z)$ is invertible for $|z|\le 1$ and $v_-(z)$ for $|z|\ge 1$.

Using symmetry relation \eqref{sym3}, we can obtain an explicit formula for $v_-(z)$. To that end, it is convenient to introduce
the functions
\begin{equation}\label{q24}
\begin{aligned}
u_+(z)=\frac{v_+(z)}{g(z)}\,,\quad u_-(z)=v_-(z)\,,
\end{aligned}
\end{equation}
where $g(z)$ is defined in \eqref{p11c}. Then
\begin{equation}\label{q25}
\begin{aligned}
u_+(z)u_-(z)=\frac{\pi(z)}{g(z)}\,, 
\end{aligned}
\end{equation}
hence by symmetry relation \eqref{sym3},
\begin{equation}\label{q26}
u_+(z^{-1})u_-(z^{-1})=[u_+(z)u_-(z)]^{-1}=u_-^{-1}(z)u_+^{-1}(z), 
\end{equation}
so that
\begin{equation}\label{q27}
\begin{aligned}
u_-(z)u_+(z^{-1})=u_+^{-1}(z)u_-^{-1}(z^{-1}). 
\end{aligned}
\end{equation}
By Liouville's theorem, there is a constant invertible matrix $C$ such that
\begin{equation}\label{q28}
\begin{aligned}
u_-(z)u_+(z^{-1})=u_+^{-1}(z)u_-^{-1}(z^{-1})=C,
\end{aligned}
\end{equation}
hence
\begin{equation}\label{q29}
u_-(z)=Cu_+^{-1}(z^{-1})
\end{equation}
and
\begin{equation}\label{q30}
u_-(z^{-1})=C^{-1}u_+^{-1}(z)\implies u_-(z)=C^{-1}u_+^{-1}(z^{-1}).
\end{equation}
Comparing \eqref{q29} with \eqref{q30}, we obtain that
\begin{equation}\label{q31}
C=C^{-1}\implies C^2=I.
\end{equation}
From \eqref{q24} and \eqref{q29} we obtain that
\begin{equation}\label{q32}
v_-(z)=u_-(z)=Cu_+^{-1}(z^{-1})=C[g^{-1}(z^{-1})v_+(z^{-1})]^{-1}=Cg(z^{-1})v_+^{-1 }(z^{-1}).
\end{equation}
Equation \eqref{q31} does not determine $C$ uniquely.
To calculate $C$ let us substitute $z=\infty$ into the latter equation. From  \eqref{q23} and \eqref{q17}
we obtain that
\begin{equation}\label{q32a}
v_-(\infty)=\tau^{-2}
\begin{pmatrix}
\di\frac{\tau}{4\eta_1} &\di \frac{i\tau^2(\eta_1+1)}{8\eta_1}\\
\di \frac{i}{2} & \di -\frac{\tau}{4}
\end{pmatrix}
\begin{pmatrix}
-\tau & 0 \\
0 & 1
\end{pmatrix}=\tau^{-1}
\begin{pmatrix}
\di -\frac{\tau}{4\eta_1} &\di \frac{i\tau(\eta_1+1)}{8\eta_1}\\
\di -\frac{i}{2} & \di -\frac{1}{4}
\end{pmatrix},
\end{equation}
and from \eqref{p11c},
\begin{equation}\label{q32b}
g(0)=-\frac{\eta_1\eta_2}{4}\,.
\end{equation}
Also, from \eqref{q20},
\begin{equation}\label{q32c}
\begin{aligned}
v_+(0)=D_0(0)P_1D_1(0)P_2D_2(0)P_3D_3(0)P_4D_4(0)
=\eta_2\begin{pmatrix}
-\eta_1 & -\di \frac{i\tau(\eta_1+1)}{2} \\
2i\eta_1 & -\tau
\end{pmatrix}.
\end{aligned}
\end{equation}
This gives 
\begin{equation}\label{q32d}
C=-\frac{4}{\tau\eta_1}\begin{pmatrix}
\di -\frac{\tau}{4\eta_1} &\di \frac{i\tau(\eta_1+1)}{8\eta_1}\\
\di -\frac{i}{2} & \di -\frac{1}{4}
\end{pmatrix}
\begin{pmatrix}
-\eta_1 & -\di \frac{i\tau(\eta_1+1)}{2} \\
2i\eta_1 & -\tau
\end{pmatrix}= I.
\end{equation}
Substituting this into \eqref{q32}, we obtain that
\begin{equation}\label{q32e}
v_-(z)=g(z^{-1})v_+^{-1 }(z^{-1}).
\end{equation}
Thus, $\pi(z)$ is factored. To finish the factorization of $\phi(z)$ it remains to factor $\sg(z)$.

From \eqref{mm6} we obtain that
\begin{equation}\label{q33}
\sg(z)=\sg_+(z)\sg_-(z)
\end{equation}
with 
\begin{equation}\label{q34}
\sg_+(z)=\frac{\tau}{(z-\tau)\sqrt{f(z)}},\quad \sg_-(z)=\sg_+(z^{-1}),
\end{equation}
where
\begin{equation}\label{q35}
f(z)=\frac{(z^2-\eta_1^2)(z^2-\eta_2^2)}{4\eta_1\eta_2}\,.
\end{equation}
Comparing $f(z)$ to $g(z)$ in \eqref{p11c}, we have that
\begin{equation}\label{q36}
g(z)=\tau^{-1}(z-\tau)f(z)\,.
\end{equation}

Summarizing our calculations, we obtain that 
\begin{equation}\label{q36a}
\phi(z)=\hat\phi_+(z)\hat\phi_-(z)
\end{equation}
with
\begin{equation}\label{q37}
\hat\phi_+(z)=\sg_+(z)v_+(z)=\frac{\tau}{(z-\tau)\sqrt{f(z)}}\,v_+(z)
\end{equation}
and 
\begin{equation}\label{q38}
\hat\phi_-(z)=\sg_-(z)v_-(z)=\frac{\tau}{(z^{-1}-\tau)\sqrt{f(z^{-1})}}\,g(z^{-1})v_+^{-1}(z^{-1})
=\sqrt{f(z^{-1})}\,v_+^{-1}(z^{-1}).
\end{equation}
Factorization \eqref{q36a} is nice but it lacks a symmetry with respect to the swap of $\eta_1$, $\eta_2$.
The reason for this is that in the decreasing power algorithm we started with factoring out $(z-\eta_1)$. See formula
\eqref{q5}. If instead we started with factoring out $(z-\eta_2)$, we would get another factorization of
$\phi(z)$, differing by a constant factor. To make the factorization symmetric, we define   
\begin{equation}\label{q38a}
\phi_+(z)=\hat\phi_+(z)P_5,\quad \phi_-(z)=P_5^{-1}\hat\phi_+(z),
\end{equation}
where $P_5$ is defined in \eqref{th6}, \eqref{th7}.

Finally, denoting
\begin{equation}\label{q39}
\Psi(z)=\frac{v_+(z)P_5}{\sqrt{f(z)}}\,,
\end{equation}
we obtain that
\begin{equation}\label{q40}
\phi_+(z)=A(z)\Psi(z),\quad A(z)=\frac{\tau}{z-\tau}\,,
\end{equation}
and
\begin{equation}\label{q41}
\phi_-(z)=\Psi^{-1}(z^{-1}).
\end{equation}
This proves Theorem \ref{factor}.

\section{Proof of Theorem \ref{abk}} \label{Appendix_C}

{\it Asymptotics of $\al_k$}. By \eqref{fdet3}, \eqref{th2} and \eqref{th10},
\begin{equation}\label{ab1}
\begin{aligned}
\al(z)=\phi_-(z)\theta_+^{-1}(z)=\frac{1}{A(z)}\,[\Psi(z^{-1})]^{-1}\sg_3[\Psi^{\rm T}(z)]^{-1}.
\end{aligned}
\end{equation}
Let us evaluate $\Psi^{-1}(z)$. By \eqref{th4},
\begin{equation}\label{ab2}
\begin{aligned}
\Psi(z)=\frac{1}{\sqrt{f(z)}}\,D(z),\quad f(z)=\frac{(z^2-\eta_1^2)(z^2-\eta_2^2)}{4\eta_1\eta_2}\,,
\end{aligned}
\end{equation}
where
\begin{equation}\label{ab3}
\begin{aligned}
D(z)=D_0(z)P_1D_1(z)P_2D_2(z)P_3D_3(z)P_4D_4(z)P_5.
\end{aligned}
\end{equation}
This gives that 
\begin{equation}\label{ab4}
\begin{aligned}
\det \Psi(z)=4\eta_1\eta_2(z-\tau)
\end{aligned}
\end{equation}
hence
\begin{equation}\label{ab5}
\begin{aligned}
\Psi^{-1}(z)=\frac{1}{\det \Psi(z)}\,\adj \Psi(z)
=\frac{1}{(z-\tau)\sqrt{4\eta_1\eta_2(z^2-\eta_1^2)(z^2-\eta_2^2)}}\,\adj D(z).
\end{aligned}
\end{equation}
Substituting this into \eqref{ab1}, we obtain that 
\begin{equation}\label{ab6}
\begin{aligned}
\al(z)&=\frac{1}{4\eta_1\eta_2\tau(z^{-1}-\tau)\sqrt{(z^{-2}-\eta_1^2)(z^{-2}-\eta_2^2)(z^2-\eta_1^2)(z^2-\eta_2^2)}}\\
&\times [\adj D(z^{-1})]\,\sg_3
\,[\adj D^{\rm T}(z)].
\end{aligned}
\end{equation}
Using identity \eqref{p9}, we can write $\al(e^{ix})$ as
\begin{equation}\label{ab7}
\begin{aligned}
\al(e^{ix})&=\frac{1}{16\eta_1^2\eta_2^2\tau(e^{-ix}-\tau)\sqrt{t^2+\sin^2 x+\sin^4 x}}\,
[\adj D(e^{-ix})]\,\sg_3
\,[\adj D^{\rm T}(e^{ix})].
\end{aligned}
\end{equation}
Observe that by  \eqref{th5}, \eqref{th6}, $D(z)$ is a quartic polynomial in $z$.

If 
$0<t<\frac{1}{2}\,$, then 
\begin{equation}\label{ab8}
\eta_1>\eta_2>1, 
\end{equation}
(see Fig.2). 
The function $\al(e^{ix})$ is analytic on the real line and $2\pi$-periodic in $x$, and by the Cauchy theorem we can move the 
contour of integration in \eqref{fd2} down to the line $\Im x=-i\ln\eta_2$:
\begin{equation}\label{ab9}
\al_k=\frac{e^{-k\ln\eta_2}}{2\pi}\int_0^{2\pi} \al(e^{i(x-i\ln\eta_2)})e^{-ikx}dx.
\end{equation}
Due to the factor $\frac{1}{\sqrt{z^2-\eta_2^2}}\,$ in \eqref{ab6}, 
the function $\al(e^{i(x-i\ln\eta_2)})$ has singularities at the points
\begin{equation}\label{ab10}
x_0=0, \quad x_1=\pi,
\end{equation}
Consider first $x_0$. For small real $x$, formula \eqref{ab6} gives that 
\begin{equation}\label{ab11}
\al(e^{i(x-i\ln\eta_2)})=\frac{c(e^{ix})}{\sqrt{1-e^{ix}}}\,,
\end{equation}
where the square root is taken on the principal sheet, with a cut on $(-\infty,0)$,
and
\begin{equation}\label{ab12}
c(z)=c_0+c_1(z-1)+\ldots
\end{equation}
is analytic at $z=1$, with
\begin{equation}\label{ab13}
\begin{aligned}
c_0&=-\frac{1}{4\eta_1\eta_2\tau(\tau\eta_2-1)\sqrt{2(\eta_1^2-\eta_2^{-2})(\eta_2^2-\eta_2^{-2})(\eta_1^2-\eta_2^2)}}\\
&\times [\adj D(\eta_2^{-1})]\,\sg_3
\,[\adj D^{\rm T}(\eta_2)]
\end{aligned}
\end{equation}
From the binomial series we have that for any $\ga>-1$,
\begin{equation}\label{ab14}
\frac{1}{2\pi}\int_0^{2\pi} (1-e^{ix})^\ga e^{-ikx}\,dx=
\frac{\Ga(k-\ga)}{\Ga(k+1)\Ga(\ga)}\,,\quad k\ge 0,\\
\end{equation}
In particular, for $\ga=-\frac{1}{2}$, as $k\to\infty$,
\begin{equation}\label{ab15}
\begin{aligned}
\frac{1}{2\pi}\int_0^{2\pi} (1-e^{ix})^{-\frac {1}{2}} e^{-ikx}\,dx&=
\frac{\Ga(k+\frac{1}{2})}{\Ga(k+1)(-2\sqrt{\pi})}\\
&\sim -\frac{1}{2\sqrt{\pi k}}\left(1-\frac{1}{8k}+\frac{1}{128k^2}+\ldots\right).
\end{aligned}
\end{equation}
This gives the contribution to $\al_k$, as $k\to\infty$, from a neighborhood of the point $x_0$ as
\begin{equation}\label{ab16}
\al_k^0\sim \frac{e^{-k\ln\eta_2}}{\sqrt k}\sum_{j=0}^\infty \frac{a_j^0}{k^j}\,,
\end{equation}
with
\begin{equation}\label{ab17}
\begin{aligned}
a_0^0&=-\frac{1}{2\sqrt{\pi}}\,c_0=
\frac{1}{8\sqrt{2\pi}\,\eta_1\eta_2\tau(\tau\eta_2-1)\sqrt{(\eta_1^2-\eta_2^{-2})(\eta_2^2-\eta_2^{-2})(\eta_1^2-\eta_2^2)}}\\
&\times [\adj D(\eta_2^{-1})]\,\sg_3
\,[\adj D^{\rm T}(\eta_2)].
\end{aligned}
\end{equation}
Similarly, the contribution to $\al_k$, as $k\to\infty$, from a neighborhood of the point $x_1$ is equal to 
\begin{equation}\label{ab18}
\al_k^1\sim \frac{(-1)^k e^{-k\ln\eta_2}}{\sqrt k}\sum_{j=0}^\infty \frac{a_j^1}{k^j}\,,
\end{equation}
with
\begin{equation}\label{ab19}
\begin{aligned}
a_0^1&=
\frac{1}{8\sqrt{2\pi}\,\eta_1\eta_2\tau(\tau\eta_2-1)\sqrt{(\eta_1^2-\eta_2^{-2})(\eta_2^2-\eta_2^{-2})(\eta_1^2-\eta_2^2)}}\\
&\times [\adj D(-\eta_2^{-1})]\,\sg_3
\,[\adj D^{\rm T}(-\eta_2)].
\end{aligned}
\end{equation}
Adding the contributions to $\al_k$ from the points $x_0$, $x_1$, we obtain that  
\begin{equation}\label{ab20}
\al_k\sim \frac{e^{-k\ln\eta_2}}{\sqrt k}\sum_{j=0}^\infty \frac{a_j^0+(-1)^k a_j^1}{k^j}\,,
\end{equation}
with
\begin{equation}\label{ab21}
\begin{aligned}
a_0^0&=\gamma_1 [\adj D(\eta_2^{-1})]\,\sg_3
\,[\adj D^{\rm T}(\eta_2),\\
a_0^1&=\gamma_1[\adj D(-\eta_2^{-1})]\,\sg_3
\,[\adj D^{\rm T}(-\eta_2)],
\end{aligned}
\end{equation}
where
\begin{equation}\label{ab22}
\begin{aligned}
\gamma_1=
\frac{1}{8\sqrt{2\pi}\,\eta_1\eta_2\tau(\tau\eta_2-1)\sqrt{(\eta_1^2-\eta_2^{-2})(\eta_2^2-\eta_2^{-2})(\eta_1^2-\eta_2^2)}}\,.
\end{aligned}
\end{equation}

{\it Asymptotics of $\be_k$}. By \eqref{fdet3}, \eqref{th2} and \eqref{th10},
\begin{equation}\label{ba1}
\begin{aligned}
\be(z)=\theta_-^{-1}(z^{-1})\phi_+(z^{-1})=[\sg_3\phi_-^{\rm T}(z^{-1})]^{-1}\phi_+(z^{-1})
= A(z^{-1})\Psi^{\rm T}(z)\sg_3 \Psi(z^{-1}),
\end{aligned}
\end{equation}
hence
\begin{equation}\label{ba2}
\begin{aligned}
\be(z)=\frac{A(z^{-1})}{\sqrt{f(z)f(z^{-1})}}\,D^{\rm T}(z)\sg_3 D(z^{-1}).
\end{aligned}
\end{equation}
This gives that
\begin{equation}\label{ba3}
\begin{aligned}
\be(z)&=\frac{4\eta_1\eta_2\tau}{(z^{-1}-\tau)\sqrt{(z^{-2}-\eta_1^2)(z^{-2}-\eta_2^2)(z^2-\eta_1^2)(z^2-\eta_2^2)}}\,
 D^{\rm T}(z)\sg_3 D(z^{-1}),
\end{aligned}
\end{equation}
which is very similar to formula \eqref{ab6} for $\al(z)$. Repeating the derivation of formula \eqref{ab20}
for $\al_k$, we obtain the asymptotic formula for $\be_k$:
\begin{equation}\label{ba4}
\be_k\sim \frac{e^{-k\ln\eta_2}}{\sqrt k}\sum_{j=0}^\infty \frac{b_j^0+(-1)^k b_j^1}{k^j}\,,
\end{equation}
with
\begin{equation}\label{ba5}
\begin{aligned}
b_0^0&=\gamma_2
\left[D(\eta_2^{-1})\,\sg_3
\, D^{\rm T}(\eta_2)\right],\\ 
b_0^1&=\gamma_2
\left[ D(-\eta_2^{-1})\,\sg_3
\, D^{\rm T}(-\eta_2)\right],
\end{aligned}
\end{equation}
where
\begin{equation}\label{ba6}
\begin{aligned}
\gamma_2=
\frac{\sqrt{2}\,\eta_1\eta_2\tau}{\sqrt{\pi}(\tau\eta_2-1)\sqrt{(\eta_1^2-\eta_2^{-2})(\eta_2^2-\eta_2^{-2})(\eta_1^2-\eta_2^2)}}.
\end{aligned}
\end{equation}

\section{Proof of Theorem \ref{abk_sup}} \label{Appendix_D}

If 
$t>\frac{1}{2}\,$, then $\eta_1$, $\eta_2$ are complex conjugate numbers,
\begin{equation}\label{app3_4}
\eta_{1}=r e^{- i\theta},\quad \eta_{2}=r e^{ i\theta};\quad 0<\theta<\frac{\pi}{4}\,. 
\end{equation} 
(See Fig.2.)
The function $\al(e^{ix})$ is analytic on the real line and $2\pi$-periodic in $x$, and by the Cauchy theorem we can move the 
contour of integration in \eqref{fd2} down to the line $\Im x=-i\ln r$:
\begin{equation}\label{app3_5}
\al_k=\frac{e^{-k\ln r}}{2\pi}\int_0^{2\pi} \al(e^{i(x-i\ln r)})e^{-ikx}dx.
\end{equation}
Due to the factor $\frac{1}{\sqrt{(z^2-\eta_1^2)(z^2-\eta_2^2)}}\,$ in \eqref{ab6}, 
the function $\al(e^{i(x-i\ln r)})$ has singularities at the four points (mod $2\pi$),
\begin{equation}\label{app3_6}
\begin{aligned}
x_{p,\sg}=(-1)^p \theta+\frac{(1-\sg)\pi}{2},\quad p=1,2,\quad \sg=\pm 1.
\end{aligned}
\end{equation}
We can use a decomposition of unity on the circle $\T=\R/(2\pi) \Z$,
\begin{equation}\label{su1}
1=\sum_{p=1,2;\,\sg=\pm 1}\chi(x;p,\sg),\quad \chi(x;p,\sg)\in C^{\infty}(\T),
\end{equation}
isolating the singular points, and  define
\begin{equation}\label{su2}
\al_k(p,\sg)=\frac{e^{-k\ln r}}{2\pi}\int_{\T} \chi(x;p,\sg)\al(e^{i(x-i\ln r)})e^{-ikx}dx.
\end{equation}
Then
\begin{equation}\label{su3}
\al_k=\sum_{p=1,2;\,\sg=\pm 1}\al_k(p,\sg),
\end{equation}
and as $k\to\infty$, $\al_k(p,\sg)$ admit the asymptotic expansions,
\begin{equation}\label{app3_7}
\al_k(p,\sg)\sim \frac{\sg^k e^{-k\ln \eta_p}}{\sqrt k}\sum_{j=0}^\infty \frac{a_j(p,\sg)}{k^j}\,,\quad p=1,2,\quad \sg=\pm 1.
\end{equation}
By \eqref{app3_6}, $x_{2,\sg}=-x_{1,\sg}$, and we may assume that 
\begin{equation}\label{app3_8}
\chi(x;2,\sg)=\chi(-x;1,\sg).
\end{equation}
Then by symmetry relation \eqref{sab4},
\begin{equation}\label{app3_9}
\begin{aligned}
\overline{\al_k(2,\sg)}&=\frac{e^{-k\ln r}}{2\pi}\int_{\T} \chi(x;2,\sg)\overline{\al(e^{i(x-i\ln r)})}e^{ikx}dx\\
&=\frac{e^{-k\ln r}}{2\pi}\int_{\T} \chi(x;1,\sg)\overline{\al(e^{i(-x-i\ln r)})}e^{-ikx}dx\\
&=\frac{e^{-k\ln r}}{2\pi}\int_{\T} \chi(x;1,\sg)\sg_3\al(e^{i(x-i\ln r)})\sg_3 e^{-ikx}dx\\
&=\sg_3\al_k(1,\sg)\sg_3.
\end{aligned}
\end{equation}
This implies that the coefficients $a_j(p,\sg)$ of asymptotic expansions \eqref{app3_7} satisfy the 
symmetry relation
\begin{equation}\label{app3_10}
a_j(2,\sg)=\sg_3 \,\overline {a_j(1,\sg)}\,\sg_3
\end{equation}

From \eqref{ab6} we obtain that the leading coefficients, with $j=0$, are equal to
\begin{equation}\label{app3_11}
\begin{aligned}
&a_0(p,\sg)=\gamma_p\,[\adj D(\sg\eta_p^{-1})]\,\sg_3\,[\adj D^{\rm T}(\sg\eta_p)],\\
&\gamma_p=\frac{1}{8\sqrt{2\pi}\,\eta_1\eta_2\tau(\tau \eta_p-1)\sqrt{(\eta_1^2-\eta_p^{-2})(\eta_2^2-\eta_p^{-2})(\eta_{p'}^2-\eta_p^2)}},
\quad p'=2-p.
\end{aligned}
\end{equation}
This proves the asymptotic expansion for the coefficients $\al_k$. Using formula \eqref{ba3}, we prove in the
same way the asymptotic expansion for the coefficients $\be_k$.

\end{document}